\documentclass[preprint1]{aastex}
\usepackage{epsfig}
\usepackage{natbib}
\usepackage{adjustbox}
\usepackage{lscape}
\begin{document}

\newcommand{\vdag}{(v)^\dagger}

\title{Herschel Observations of Dusty Debris Disks}

\author{Laura Vican$^{1}$ \&  Adam Schneider$^{2}$ \& Geoff Bryden$^{3}$ \& Carl Melis$^{4}$ \& B. Zuckerman$^{1}$  \& Joseph Rhee$^{5}$ \& Inseok Song$^{6}$}   

\affil{$^{1}$ Department of Physics and Astronomy, University of California, 
Los Angeles, CA 90095, USA}
\affil{$^{2}$ Department of Physics and Astronomy, University of Toledo, 
Toledo, OH 43606, USA}
\affil{$^{3}$ JPL/Caltech}
\affil{$^{4}$ Center for Astrophysics and Space Sciences, University of California, San Diego, CA 92093-0424, USA}
\affil{$^{5}$ California State Polytechnic University}
\affil{$^{6}$ Department of Physics and Astronomy, University of Georgia, 
Athens, GA 30602-2451, USA}

\begin{abstract}
We present results from two Herschel observing programs using the Photodetector Array Camera and Spectrometer. During three separate campaigns, we obtained Herschel data for 24 stars at 70, 100, and 160 $\mu$m. We chose stars that were already known or suspected to have circumstellar dust based on excess infrared emission previously measured with IRAS or Spitzer, and used Herschel to examine long-wavelength properties of the dust. Fifteen stars were found to be uncontaminated by background sources, and possess infrared emission most likely due to a circumstellar debris disk. We analyzed the properties of these debris disks to better understand the physical mechanisms responsible for dust production and removal. Seven targets were spatially resolved in the Herschel images. Based on fits to their spectral energy distributions, nine disks appear to have two temperature components. Of these nine, in three cases, the warmer dust component is likely the result of a transient process rather than a steady state collisional cascade. The dust belts at four stars are likely stirred by an unseen planet, and merit further investigation.
\end{abstract}

\section{Introduction}
Debris disks are signposts of planetesimal formation and, as such, are crucial subjects for study when considering the evolution of a planetary system. Debris disks are typically identified via an excess infrared (IR) flux above the stellar photosphere. New debris disks have been discovered with five satellites, starting with the \textit{InfraRed Astronomical Satellite (IRAS)} discovery of the first debris disk around Vega in 1984 \citep{Aumann_1984}. In 1995, the Infrared Space Observatory (ISO; \citealt{Boulade_1995}) imaged the sky at wavelengths ranging from 2.5 to 240 $\mu$m, providing low-resolution long wavelength photometry. Subsequently, the \textit{Spitzer Space Telescope} \citep{Werner_2004} provided mid-IR spectroscopy with the Infrared Spectrograph (IRS; \citealt{Houck_2004}), and photometry with the Multiband Imaging Photometer (MIPS; \citealt{Rieke_2004}), leading to the detection of over 100 new debris disks (e.g. \citealt{Zuckerman_2004}, \citealt{Plavchan_2009}, \citealt{Chen_2009}, \citealt{Chen_2014}). The Wide Field Survey Explorer (WISE; \citealt{Wright_2010}) imaged the sky at 3.4, 4.6, 11, and 22$\mu$m, and was also most sensitive to warm and hot debris disks that peak in the mid-IR.

In 2009, the \textit{Herschel Space Observatory} \citep{Pilbratt_2010} began taking data, providing far-IR photometry with the Photodetector Array Camera and Spectrometer (PACS; \citealt{Poglitsch_2008}) and spectroscopy with the Spectral and Photometric Imaging Receiver (SPIRE; \citealt{Griffin_2008}) and the Heterodyne Instrument for the Far Infrared (HIFI; \citealt{deGraauw_2008}). Where Spitzer was uniquely able to detect warm ($>$100 K) debris disks in the terrestrial planet zone, Herschel was sensitive to cooler disks ($<$100 K) at larger radial separations from their host stars. In this paper, we present Herschel observations of 24 stars initially identified with IRAS and/or Spitzer as definitely or possibly possessing a debris disk.

One goal of our project was to search for cold dust components that would peak near the Herschel PACS wavelengths (70, 100, and 160 $\mu$m). In the case that there is no separate cold dust component present, Herschel photometry helps to better characterize the Rayleigh-Jeans tail of thermal emission from warm dust.

Another goal was the identification of disks with double-belt debris systems; that is, systems containing an inner belt of warm or hot ($>$100K) dust and an outer belt of cold ($<$100 K) dust. Such systems would be a direct analog of our own Solar System, which hosts a Kuiper Belt that lies between 30-50 AU at $\sim$50 K and an Asteroid Belt at 3 AU and $\sim$175 K. A double-belt system may also be a signature of a planet (or planets) that lie in the gap between dust belts. Such systems have been discovered around HR 8799 (\citealt{Marois_2008}, \citealt{Marois_2010}, \citealt{Matthews_2010}) and HD 95086 (\citealt{Rameau_2013}, \citealt{Su_2015}).

\section{Stellar Sample}
The sample used in this work is an amalgamation of stellar samples from three different Herschel proposals (bzuckerman-OT1, jolofsson-OT1, and bzuckerman-OT2). In the first OT1 proposal (PI: B. Zuckerman), we observed A-F stars with known, very luminous mid-IR emission. Since mid-IR emission is known to originate from the terrestrial planet region, we wanted to use Herschel to search for accompanying cold dust in outer regions analogous to the Solar System's Kuiper Belt. Four such stars were observed during this program. The second OT-1 proposal (PI: J. Olofsson) had very similar science goals, in that the authors were looking for cold dust components accompanying known warm debris disks. Six stars were observed for their program, of which we present two which fit into our initial selection criteria (luminous mid-IR emission).

The objective of the OT-2 proposal was to follow up on a subsample of stars (from \citealt{Rhee_2007} and \citealt{Zuckerman_2011}) that had been observed either with IRAS or Spitzer, and that had only one far-IR data point at 60 $\mu$m (IRAS) or 70 $\mu$m (Spitzer). Herschel observations were carried out to characterize the far-IR emission of the dust and, in the case of stars with apparent excess emission detected only at 60 $\mu$m with IRAS, to confirm or deny the existence of a dusty debris disk. IRAS is not only less sensitive than Herschel, but its large beam size made IRAS vulnerable to confusion by background sources (the Herschel PSF is 5.75$^{\prime\prime}$ at 70 $\mu$m, while the average detector element size for IRAS at 60 $\mu$m was 5$^{\prime}$ $\times$ 2$^{\prime}$). Eighteen stars were observed during this OT-2 program. We collected information from the literature on binarity, distance from Earth, and stellar age. Literature data for target stars can be found in Table \ref{table:stellar}. 
 
\section{Observations}
We observed 24 stars in total with PACS; simultaneous observations were obtained at 160 $\mu$m and either 70 $\mu$m or 100 $\mu$m. Although some stars were not detected at 160 $\mu$m, a 3$\sigma$ upper limit at that wavelength helped to constrain cold dust temperatures. The observations were taken with a scan speed of 20$\arcsec$/sec.

We used HIPE (Herschel Interactive Processing Environment, version 12.0; \citealt{2010ASPC..434..139O}) to reduce the data and produce the final maps. We used a pixel scale of 1$\arcsec$/pixel for 70 and 100 $\mu$m data and 2$\arcsec$/pixel for the 160 $\mu$m data. A high pass filter was applied to remove instrumental noise. To achieve the highest signal to noise in the resulting maps, we used a high pass filter radius of 30$\arcsec$ for 70$\mu$m data and 70$\arcsec$ for 100 and 160$\mu$m data. The larger radius filter was used for the longer wavelength data in order to be more aggressive in removing instrumental noise at those wavelengths while maintaining a high SNR. A mask was applied to avoid removing any flux within 15" of the target star.

Aperture photometry was carried out with an aperture radius of 5$\arcsec$ for unresolved sources and 10$\arcsec$ for resolved sources, with a sky annulus extending from 20-40$\arcsec$. Aperture corrections were made to account for photospheric flux falling outside of the chosen aperture$\footnotemark{}$$\footnotetext{All corrections were taken from Herschel Release Note PICC-ME-TN-037, Table 15.}$. Errors on the photometric points are derived by placing apertures on empty regions of sky along the sky background annulus and measuring the r.m.s. of the background flux. Some targets presented flux at a significant offset from the stellar location. If the offset was larger than the pointing accuracy of Herschel ($\sim$2$\arcsec$), we excluded it from further analysis. This was the case for two stars in our target list (HD 60234 and HD 203562). These two stars were taken from \citet{Rhee_2007}, and were 60$\mu$m excesses only. Observed fluxes are found in Table 2. We discuss potential sources of contamination in Section \ref{contamination}.

\subsection{Herschel Non-Detections}
A non-detection with Herschel implies that any circumstellar dust would necessarily be $<$35K, extremely tenuous (L$_{IR}$/L$_{bol}$=$\tau$$<$10$^{-6}$), or both. The three stars mentioned in this section were not detected in any of the Herschel bands, and are not considered in the analysis in later sections.

\subsubsection{HD 70298}
This star was reported by \citet{Rhee_2007} as a new debris disk candidate with a surprisingly substantial 60$\mu$m excess  ($\tau$=3.54E-04), given its old age ($>$3Gyr). The star was not observed by Spitzer, but a WISE excess was identified by \citet{McDonald_2012} at 22 $\mu$m ($\tau$=4.35E-04). Close inspection of the WISE images shows that there is a nebulous IR source $\sim$15$\arcsec$ away from the target, which is inside the WISE contamination radius at 22 $\mu$m (2 $\times$ FWHM = 24$\arcsec$; \citealt{Cutri_2012}). We therefore consider the WISE and IRAS images to be contaminated. In the absence of any detectable flux at Herschel wavelengths, we report no IR excess around this star. 

\subsubsection{HD 72660}
This star was reported by \citet{Rhee_2007} as a new debris disk candidate based on an IRAS excess. The star was subsequently observed with Spitzer, and a mid-IR spectrum showed no IR excess consistent with the IRAS data point. WISE also saw no evidence of any warm excess in the near-IR. Even so, it remained possible that the IRAS photometry was catching the Wein tail of cold dust ($<$50 K), until our negative Herschel observations. 

\subsubsection{HD 132950}
This star was reported by \citet{Rhee_2007} as a new debris disk candidate with a large IR excess ($\tau$=1.17E-03) and an old age ($\sim$3 Gyr). A Spitzer IR spectrum shows no evidence of an IR excess consistent with the IRAS data point, and no warm excess was seen by WISE. In the absence of an IR excess seen with Herschel, we report no detectable dust around this star. 

\subsection{Systems with No Detectable Dust}
In addition to stars with no detectable flux at Herschel wavelengths, there is one system (HD 191692) for which Herschel detected the photosphere of the star, but no evidence of an IR excess was seen. This star was reported by \citet{Rhee_2007} to have a small ($\tau$$<$10$^{-5}$) IR excess seen at IRAS wavelengths, but unconfirmed by Spitzer. Their dust model fit to the IRAS photometry suggested the existence of dust at over 200 AU separation from the central star. \citet{Rodriguez_2012}  reported the star as a binary with a separation of $<$1 AU, meaning that the dust, if confirmed, would be circumbinary. \citet{McDonald_2012} report a large IR excess ($\tau$$>$10$^{-3}$) seen at WISE wavelengths. A close inspection of the WISE images shows a source $\sim$18$\arcsec$ to the NE of the target star, which is within the contamination radius of the WISE beam at 22$\mu$m. Thus we consider both the IRAS and the WISE images to be contaminated, and given that the Herschel images show no evidence of an IR excess, we report this star to have no detectable dust. 

\section{Sources of Possible Contamination} \label{contamination}
Of the remaining targets (those with detectable Herschel fluxes indicative of an IR excess), four (HD 8558, HD 13183, HD 80425, and HD 99945) have apparent cold dust disks with temperatures $<$40K (see Section \ref{SEDs}). To ensure that we are really seeing evidence of cold debris belts, we investigated several alternative sources of the apparent IR excess.

\subsection{Extragalactic Background}
\citet{Gaspar_2014} presented Herschel observations of cold debris disks and investigated the possibility of confusion with a background galaxy. A typical galaxy below the confusion limit of Herschel ($\sim$2.5mJy at 160 $\mu$m) would lie between z=0.94 and 1.2 and its IR emission would correspond to an apparent dust temperature which would peak between 20 and 29 K \citep{Magnelli_2013}. Of the four stars with Herschel emission at 160$\mu$m, only two (HD 8558 and HD 80425) have F$_{160}$ $<$ 10 mJy and could be explained by confusion with a background galaxy at z$\sim$1. In addition, the position of a third star (HD 13183) is 2$\arcsec$ away from a galaxy detected with GALEX. Since the galaxy is within the confusion beam radius of PACS at 100$\mu$m (7.19$\arcsec$; \citealt{Gaspar_2014}), we consider this target star to be contaminated, even though the source flux is 40mJy at 160 $\mu$m.  These three stars are not included in Table \ref{dustpars}.

\subsection{IR Cirrus}
We must also investigate the possibility that excess IR emission is due to confusion with background cirrus, which is known to emit at $\sim$20 K \citep{Roy_2010}. We checked each of our cold excess systems (HD 8558, HD 13183, HD 80425, and HD 99945) for evidence of cirrus in the Herschel and WISE images. HD 99945 is not explained by confusion with a background galaxy or nearby IR excess source. We also found no detectable cirrus in the Herschel or WISE images. This, combined with the fact that the 160 $\mu$m flux for this object is $>$100mJy implies that the excess around the star is due to emission from a debris disk. HD 99945 is therefore included in Table 3. Of the four potential cold disks in our sample, only HD 99945 (the warmest of the four) survived to the final sample.

\section{Spectral Energy Distributions}\label{SEDs}
Spectral energy distributions (SEDs) were created using a fully automated photosphere-fitting technique that made use of the PHOENIX models \citep{Hauschildt_1999}. Stellar photospheres were fit using B, V, J, H, and K fluxes, while the mid- and far-IR photometry was used to fit the dust emission. We fit a simple blackbody to the IR photometry using Spitzer, WISE, and IRAS to supplement the Herschel data points whenever possible. For some systems, the long wavelength data (at 100 or 160 $\mu$m) falls below a simple blackbody curve. In these cases, we applied a modified blackbody described by:

\begin{equation}
F_{\nu}\propto\left(\frac{\nu_{0}}{\nu}\right)^{\beta}B_{\nu}(T_{d})
\end{equation}

\noindent where T$_{d}$ is the dust temperature. Uncertainties for T$_{d}$ are found using a Monte Carlo approach. For each SED, we generated 1000 simulated data sets by randomly drawing flux values from a single Gaussian distribution centered on the observed fluxes, where the width of the distribution is given by the observed uncertainty.  The values and uncertainties for T$_{d}$ reported in Table 3 are the average and standard deviation of the ensemble of fits for each object. The resulting SEDs are found in Figure \ref{fig:SEDs1}.

\subsection{Stellar Parameters}
We used the best-fit model photospheres to determine the stellar temperature and radius. From there, we calculated a stellar luminosity using L$_{*}$=4$\pi$R$_{*}$$^{2}$$\sigma$T$_{eff}$$^{4}$. We assumed solar metallicity and log(g) for all 24 stars. Stellar masses were taken from the literature. Ages for stars in our sample were taken from moving group membership whenever possible. When moving group membership could not be determined, we used literature ages from stellar isochrones (for A-F-type stars). For all other stars, we relied on literature ages based on lithium abundance or X-ray activity.  Stellar parameters are found in Table \ref{table:stellar}. 

\subsection{Dust Parameters}
For the 15 stars in our sample with dust detected by Herschel, we obtain a dust temperature and fractional IR luminosity (L$_{IR}$/L$_{bol}$=$\tau$) from a blackbody SED fit. Assuming blackbody dust grains, the orbital semi-major axis is:

\begin{equation}
R_{BB}=\frac{R_{*}}{2}\left(\frac{T_{*}}{T_{BB}}\right)^{2}
\end{equation}

For the purposes of this work, we define three dust temperature regions; hot dust ($>$ 200K), warm dust (100-200K), and cool dust ($<$ 100K). These temperature regions were chosen to correspond to the solar system's zodiacal dust, asteroid belt, and Kuiper belt. Of the 15 systems in our sample that were detected by Herschel and were found to be uncontaminated, 9 stars show a cool dust component, 3 stars show a warm dust component, and 10 stars show a hot dust component (some stars show multiple dust components). 

It should be noted that the dust temperatures provided in Table \ref{dustpars} are the best-fit values from SED fitting described in Section \ref{SEDs}. In some cases, the best fit dust temperature is reported as being $>$1000 K. We examined the variation in $\chi$$^{2}$ values for our SED fits as we varied the dust temperatures between 600-1200K, and find that the $\chi$$^{2}$ values drop significantly as one increases the dust temperature from 600 to 1000 K, but flatten out after 1000 K. Such model-suggested dust temperatures of $\gtrsim$1000 K are difficult to explain (especially since the hottest dust temperature around an extremely dusty star known to this point is the $\sim$800 K dust around V488 Per, reported by \citealt{Zuckerman_2012b}) and we acknowledge that a one or even two ring model for these debris disks is very likely an over-simplification of the actual dust configuration.

We calculated the blowout radius for a dust grain, a$_{blow}$, defined as:

\begin{equation}
a_{blow}=\frac{3L_{*}Q_{PR}}{16\pi GM_{*}c\rho}
\end{equation}

where Q$_{PR}$ is the radiation pressure coupling coefficient and $\rho$ is the density of a typical dust particle \citep{Chen_2001}. We assume Q$_{PR}$ $\sim$1 for 2$\pi$a/$\lambda$ $>$ 1 and $\rho$$\sim$2.5 g cm$^{-3}$.

Finally, we calculated the minimum mass of dust in the disk. This parameter is best measured using far-IR or sub-mm data, which probes the largest grains in the system. The large grains are where most of the mass of the dust is concentrated. However, we can calculate a minimum dust mass from Equation 4 from \citet{Chen_2001}:

\begin{equation}
M_{d, min}\geq\frac{16}{3}\pi\tau\rho R_{d}^{2}\langle a\rangle\
\end{equation}

where $\langle$a$\rangle$ = (5/3)a$_{blow}$ \citep{Chen_2001} and R$_{d}$ is the dust semi-major axis. Minimum dust mass values are calculated using the outermost dust belt parameters only, since the mass of the colder dust dominates the system in all cases. 

\section{Resolved Disks}\label{resolved}

To determine whether a disk was resolved, we compared the radial profile of the star+disk to a reference PSF (see Figure \ref{fig:profs}). To create the radial profile, we binned the pixels of each raw image by radius. The error bars in Figure \ref{fig:profs} are the standard deviation of the fluxes within each radial bin. We consider each disk presented in Figure \ref{fig:profs} to be resolved$\footnotemark{}$$\footnotetext{\scriptsize HD 121191 is marginally resolved, and further discussion can be found in Section \ref{HD121191}.}$.

We performed a PSF subtraction and modeled the residual flux with a single narrow ring of dust ($\Delta$R$_{d}$=0.1R$_{d}$). The bright star $\alpha$ Cet was used as a reference PSF. The free parameters fit in our ring model were the semi-major axis, inclination, and position angle of the narrow ring. This model (convolved with the instrument PSF) was appropriate for all of our resolved systems, since residual maps showed no significant structure once the ring model was subtracted. Raw images, PSF-subtracted images, and residuals can be found in Figure \ref{fig:mapsA}. Of the 15 stars in our sample, seven (HD 54341, HD 76543, HD 76582, HD 84870, HD 85672, HD 99945, and HD 121191) were resolved at either 70 or 100 $\mu$m. One system (HD 76582) was  resolved at both 100 and 160 $\mu$m. Disk parameters determined by ring-fitting can be found in Table \ref{resolvedtable}. Errors were determined by varying parameters until residuals increased by 1$\sigma$.

\subsection{Disk Radii}
The blackbody disk semi-major axes derived from SED fitting (R$_{BB}$) often did not agree with those observed in the resolved images (R$_{img}$). We define a semi-major axis ratio  f$_{R}$=R$_{img}$/R$_{BB}$. It is well-known that small blackbody particles tend to be super-thermal and thus blackbody SED-fitting will underestimate radial extent of the dust, often by a factor of 2-5 \citep{Rodriguez_2012}. The steeper the size distribution of the dust grains, the more small grains are present in the disk, and the disk will appear much hotter than a blackbody. At first glance, then, it would seem that f$_{R}$ can probe the size distribution of the dust. 

If, however, the dust production mechanism is similar among all of our disks (see Sections 6-7), then it is reasonable to assume that the resulting size distribution would be similar as well. If that is true, the biggest factor affecting f$_{R}$ would be the spectral type or luminosity of the host star. If the host star is very luminous, it will remove the smallest grains from the system via radiation pressure; this will make the dust appear to act more like a blackbody.  

We compared our radius ratios to those of similar programs (\citealt{Morales_2013}, \citealt{Booth_2013}, \citealt{Rodriguez_2012}) in Figure \ref{fig:radii}. We find that, in most cases,  f$_{R}$ increases for lower-luminosity stars. One interesting feature to note in Figure \ref{fig:radii} is that the value of f$_{R}$ (=R$_{img}$/R$_{BB}$) is well constrained for disks that are resolved in thermal emission, but there is much more scatter in the relationship between f$_{R}$ and L$_{bol}$ for stars whose ``resolved" radii are derived from SED modeling of the Si emission (see Figure \ref{fig:radii}) and/or scattered light. This may be because the scattered light images are most sensitive to the smallest grains, which can be found in a large, extended disk or halo, rather than in a narrow ring as described by a blackbody fit. 

%\begin{bfseries}
It appears that the stars in our sample (red data points in Figure \ref{fig:radii}) have a higher scatter in L$_{bol}$- f$_{R}$ space than do stars from \citet{Booth_2013}. Notably, the biggest outliers (HD 121191 and HD 85672) are those stars for which we believe an unseen planet could be responsible for dust production (see Section \ref{planet_stirred}). Since the ratio f$_{R}$ depends heavily on the size of the grains themselves, it is possible that a planet-stirred disk has an inherently different grain size distribution than a self-stirred disk. Different grain size distribution could arise from a number of factors, including the collisional velocities of the $<$100km bodies that produce the dust, and the composition of the dust itself. It is also possible that the Herschel images are sensitive to small grains further from the star (which would peak at Herschel wavelengths), while the average temperature of the grains in the system is actually higher than the temperature implied by the blackbody fit to the SED.
%\end{bfseries}

\section{Dust Production and Planet Formation}
Having determined dust properties from the Herschel photometry, we turn our attention to the way in which the observed dust was produced. By the time a star is $\sim$10 Myr old, any primordial dust should have been completely depleted (used to form planetesimals, accreted onto the star, or blown out of the system by radiation pressure).$\footnotemark{}$ $\footnotetext{\scriptsize This idea has been challenged by recent discoveries of 30-40 Myr old disks with copious molecular gas present, suggesting that, for some stars, the protoplanetary stage may last longer than previously thought (e.g. \citealt{Kospal_2013}, \citealt{Zuckerman_2012}). Still, since there are so few stars that exhibit gas at 30-40Myr, we maintain that most primordial gas and dust should be gone by $\sim$10 Myr.}$ Thus, the dust we see in debris disks older than $\sim$10 Myr is likely second-generation, created in collisional processes. Furthermore, the lifetime of this dust against radiation pressure and drag forces is shorter than the lifetime of the star. This means that any dust seen in these systems must be replenished regularly. 

There are various mechanisms by which dust can be produced as a planetary system forms and evolves. First, the dust can be produced through steady-state collisions of small planetesimals \citep{Dominik_2003}.  This occurs naturally as a final stage in the process of planet formation. Planet formation proceeds first through the growth of $\sim$1 km-sized planetesimals. Once the planetesimals grow large enough to start gravitationally focusing the primordial dust in their path, runaway growth can occur. Following this, a few large bodies accrete a large amount of dust, and as they grow, they become the dominant accreter in their orbital path - a process known as oligarchic growth. Once $\sim$1000 km-sized bodies have formed, they can begin to dynamically stir the remaining smaller ($<$100 km) planetesimals, driving them to collisional velocities high enough to be destructive. It is in these destructive collisions that the observed dust is produced. The resulting debris is collisionally ground down until the grains are small enough to be blown out of the system via radiation pressure. On their way out of the system, they can also collide with small planetestimals, creating an outward-moving collisional cascade. 

In the process described above, the stirring mechanism at work is the natural formation of 1000 km-sized bodies in the disk itself (hence, the disks are ``self-stirred", see Section \ref{selfstirred}). However, one can also imagine a situation where small ($<$100km) bodies are stirred by a nearby planet (see Section \ref{planet_stirred}). In either case, one would expect moderate fractional IR luminosities ($\tau$$\sim$10$^{-4}$) compared to those associated with protoplanetary disks ($\tau$$\sim$10$^{-3}$).  

Alternatively, a belt of dust can be created through a giant impact between planetary embryos (e.g. \cite{Jackson_2012}; see Section \ref{giantimpacts}). In an even more extreme scenario, two fully formed rocky planets can undergo a catastrophic collision, resulting in large amounts of debris. Such a collision has been proposed to explain warm dust in orbit around BD +20 307 (\citealt{Song_2005}, \citealt{Zuckerman_2008}). Such catastrophic collisions should produce a multitude of small grains. Similar collisions are also more likely to occur in the terrestrial planet formation zone, where disk surface density is high; such collisions result in production of warm or hot dust.

A third possible source of hot dust in disk systems might be debris left behind by so-called ``star-grazing comets." This mechanism was suggested by \citet{Morales_2011} who found that hot dust in a sample of Spitzer-observed systems had a characteristic temperature of 190K - near the sublimation temperature for icy bodies such as comets (see Section \ref{comets}). 

\subsection{Distinguishing Transient from Steady-State Events}\label{distinguishing}
To distinguish dust created in a transient event from dust created in a steady-state process, one can examine the maximum fractional IR luminosity that can be attained through a steady-state process ($\tau$$_{max}$). This method was explored in \citet{Wyatt_2007}. We examine the $\tau$$_{max}$ method of \cite{Wyatt_2007}. According to their Equation 20:

\begin{equation}
\tau_{max}=0.58 \times 10^{-9} R_{d}^{7/3}\frac{\Delta R_{d}}{R_{d}}D^{1/2}Q^{5/6}e^{-5/3}M_{*}^{-5/6}L_{*}^{-1/2}t^{-1}
\end{equation}

As in Equation 4, R$_{d}$ is the semi-major axis of the dust, and here $\Delta$R$_{d}$ is the radial width of the disk, D is the diameter of the largest planetesimal in the disk, Q is the specific incident energy required to destroy a particle, e is the eccentricity of the dust ring, and t is the stellar age. For our calculations, we use the outermost dust component (in cases where two components are needed to fit the observed IR flux). In all following calculations, if the dust was spatially resolved, we use the dust semi-major axis from the ring-fitting algorithm (R$_{img}$, see Section 6.1). Otherwise, we use R$_{BB}$ from the SED fits. To model a steady-state collisional cascade, we follow the prescription of \citet{Wyatt_2007} and assume that D=2000km, Q=200 J kg$^{-1}$, e=0.05, and $\Delta$R$_{d}$=0.1R$_{d}$. According to Wyatt et al., one should expect that observed dust was likely created in a transient event if $\tau$/$\tau$$_{max}$$>$1000. The $\tau$/$\tau$$_{max}$$>$1000 threshold is based on observations of known very luminous debris disks, while taking into consideration the assumptions that went into Equation 5. For the purposes of our analysis, we assume that debris disks with $\tau$/$\tau$$_{max}$$>$100 are likely to be transient in nature, while those with $\tau$/$\tau$$_{max}$$>$1000 cannot be explained by a steady-state collisional process alone (see Tables 3 and 5). We calculated $\tau$$_{max}$ for the stars in our sample and found that three debris disks systems could not be explained by a steady-state process alone (all from OT1, and thus with the highest fractional IR luminosities and previously suspected to be the result of transient processes.); HD 15407 \citep{Melis_2010}, HD 113766 \citep{Lisse_2008}, and BD +20 307 \citep{Song_2005}. Since these are well-studied systems, we do not discuss their transient nature further in this paper. Furthermore, the disks at HD 23514 and HD124718 had 100 $<$$\tau$/$\tau$$_{max}$$<$1000, implying transient processes.

\subsection{Steady-State Collisions - Stirring Mechanisms}
If the observed dust is produced in a steady-state collisional cascade, we would like to identify the trigger mechanism that starts the cascade. We consider collisions triggered by perturbations due to 1000 km-sized bodies within the disk itself (self-stirring), and collisions triggered by a nearby planet or distant stellar or substellar companion (planet-stirring). 

\subsubsection{Self-Stirring}\label{selfstirred}
In the case of self-stirring, 1000 km-sized bodies naturally form in the disk which dynamically stir the population of smaller ($<$100 km-sized) bodies to velocities high enough to cause destructive collisions. These initial collisions then trigger a collisional cascade which leads to the production of copious amounts of dust \citep{Wyatt_2008}.

Since the self-stirring mechanism requires the existence of 1000 km-sized bodies at approximately the same radial location of the dust, we can place limits on the parameters of a disk stirred by this mechanism \citep{Moor_2015}. In Figure \ref{fig:1000km}, we examine the semi-major axis of 1000 km-sized bodies as a function of stellar age. Following the models of \citet{Kenyon_2008}, we assume that 1000 km-sized planetesimals form at a time:

\begin{equation}
t_{1000}=145x_{m}^{1.15}\left(\frac{R_d}{80 AU}\right)^{3}\left(\frac{2M_{\sun}}{M_{*}}\right)^{3/2} [Myr]
\end{equation}

As in \citet{Moor_2015}, we vary two parameters in our models; the mass of the host star and a scaling factor (x$_{m}$) related to the initial mass of the protoplanetary disk (where x$_{m}$=1 corresponds to the minimum mass solar nebula).

We plot several models for the star-disk system in Figure \ref{fig:1000km}, representing stars from 1.5-2.5M$_{\sun}$ and a range of x$_{m}$. For self-stirring to be responsible for the dust observed around the four labeled stars, the initial disk would need to be 30 times as massive as the minimum mass solar nebula or more. Since $\sim$30 $\times$ MMSN is approximately the protoplanetary disk mass of a high-mass star, we take this as an upper limit for our models \citep{Williams_2011}. None of the stars in our sample have detected planets.

\subsubsection{Planet Stirring}\label{planet_stirred}
For disks that are unlikely to have formed 1000 km-sized bodies at the radial location of the observed dust, planet-stirring offers another mechanism for dust production (major planets could have formed near the dust at an earlier time). Since debris disks are often considered evidence of planet formation \citep{Zuckerman_2004}, one might expect a correlation between the existence of planets and detection of circumstellar dust. Unfortunately, with so few stars observed to have both a dust disk and at least one planet, this relationship has proved difficult to study. \citet{Bryden_2009} found little correlation between planet hosts and detection of an IR excess (the typical age of stars in their sample was $\sim$6 Gyr, whereas the stars in our sample are significantly younger). More recently, \citet{Moro_2015} examined a large ($>$200 star) sample of Herschel-observed stars to look for correlations between the presence of a debris disk and: (1) the presence of low-mass planets, (2) the presence of high-mass planets, (3) metallicity, and (3) the presence of one or more stellar companions. Even with their large sample size, \citet{Moro_2015} found no significant correlations between any of the aforementioned parameters. 

In our sample of 24 stars, none have detected planets (exoplanets.org; \citealt{exoplanets}). One can try to draw conclusions about the properties of a hypothetical unseen planet based on the observed dust properties. Unfortunately, it is difficult to resolve the degeneracy between planet mass, orbital eccentricity, and semi-major axis using simple models. This issue is treated in more detail in a theory paper motivated by the present results \citep{Nesvold_2016}. 

It is possible that a belt of small planetesimals could be stirred up to collisional velocities by a distant stellar-mass companion (or brown dwarf) as opposed to a nearby planet \citep{Zuckerman_2015}. However, \citet{Rodriguez_2015} found that stars with debris disks are less likely to be found in binary systems, at least in a sample of FGK stars. This may be attributable to the fact that a companion will accelerate the evolution of the dust, making a debris disk detectable for a shorter period of time, earlier on in the evolution of the system. In any case, we do not consider companion stirring to be the cause of the disks in our sample. This possibility is also examined further in \citet{Nesvold_2016}. 

\subsection{Star-Grazing Comets}\label{comets}
One suggested explanation for the existence of warm and hot dust disks is that so-called ``star-grazing comets" leave behind a cloud of debris \citep{Morales_2011}. Should this be the case, one would expect most of this cometary debris to be at a radial separation from the star such that the dust is at a temperature of $\sim$150 K (the sublimation temperature for icy planetesimals). In Figure \ref{fig:T_hist}, we compare the temperature of our disks to those of \citet{Chen_2014} (Spitzer Catalog of Debris Disks). \citet{Morales_2011} found that, in a sample of $\sim$70 stars, most of the warm dust components fell around $\sim$190 K.$\footnotemark{}$ $\footnotetext{\scriptsize The reason the dust temperature is higher than the sublimation temperature is that once the comets sublimate and leave behind grains, those grains appear to be hotter than the original comet because they are not emitting like blackbodies.}$ We do not find a similar trend in the Spitzer catalog taken as a whole. 

The disks in our Herschel sample are mostly cold ($<$100 K), peaking around 60 K. In fact, the Spitzer data from \citet{Chen_2014} also shows a peak around 60K. A similar peak is found in the sample of \citet{Morales_2011}, but no explanation was put forward.  Perhaps this 60 K peak is simply an observational bias (a 60K dust belt would peak between 70-100 $\mu$m, close to both Spitzer and Herschel filter wavelengths). It may also be that disk detectability falls off at lower temperatures, producing a false ``peak" near 60K. \citet{Ballering_2013} also identified a peak at 60 K in a Spitzer sample of debris disks, but claim that the peak is not solely due to the dust temperature (see their Figure 5). They suggest that the cold dust temperatures follow a trend with spectral type; hotter, earlier-type stars have warmer outer debris belts. This implies that the peak at 60 K is not due to a temperature-dependent phenomenon such as sublimation.

Regardless of the exact distribution of temperatures among our debris disks, we do not find a peak in dust temperature near the sublimation temperature for comets, and thus conclude that debris left by star-grazing comets does not contribute significantly to the dust seen in the present sample of stars or in the Spitzer Catalog of Debris Disks.

\subsection{Giant Impacts and Catastrophic Collisions}\label{giantimpacts}
It is possible that the observed warm dust components are created during a giant impact between two planetary embryos in the terrestrial planet formation zone \citep{Kenyon_2005}. It is generally believed that such impacts are an important part of our own Solar System's history; a collision between the early Earth and a Mars-sized planetary embryo was likely responsible for the formation of our Moon \citep{Hartmann_1975}. We also have evidence of large catastrophic collisions around stars other than our Sun; such is the case with BD+20 307 (\citealt{Song_2005}, \citealt{Zuckerman_2008}). Even if the observed dust luminosity is smaller than $\tau$$_{max}$ (see Section \ref{distinguishing}), the dust could still have been created in a catastrophic collision, but one in which the parent bodies did not entirely pulverize one another, and much of the mass of the parent body survived in large fragments. 

\section{Two-Temperature Systems}
To determine whether the debris disks in our sample require a two-component fit, we examined the $\chi$$^{2}$ values for a single- and double- belt fit to the SEDs. In about half of the targets, we found that a double-belt fit resulted in a significantly lower $\chi$$^{2}$. In one case (HD 76543), we do not find any significant difference in the $\chi$$^{2}$ value for a single versus double belt fit, but decided to fit two belts in order to be consistent with previous studies in the literature.

Nine stars in our sample showed evidence of a two-temperature component dust system based on the SED fits. While IRS spectra were not available in all cases, which limits our ability to properly judge whether two temperature components are truly needed, we based our SED fits on a $\chi$$^{2}$ analysis, with the understanding that our results would be improved by mid-IR spectroscopy. In most cases, we would interpret a two-temperature component fit as two spatially separated disks. Such systems can be explained in several ways; 1) the dust originated in a cold belt, and "leaked" into the warm belt via PR drag and/or scattering by large planetesimals, 2) there was once one extended disk, into which a gap was carved by intervening planets, 3) the dust may have originated in two belts independent of one another, analogous to the asteroid and Kuiper belts, or 4) there is only one belt of dust that presents as a multiple temperature component system due to a range of grain properties. 

Concerning possibility 4$\#$ above, a two temperature SED could be necessary even if the dust emission arises from a single narrow belt if the grains in the belt have multiple sizes \citep{Kennedy_2014}. Since smaller grains radiate inefficiently at long wavelengths, they retain heat and appear to be at a hotter temperature than larger grains at the same radial location from the star.

Four of the two temperature systems orbit A-type stars (HD 54341, HD 76543, HD 121191, and HD 131488). Since the blowout radius of grains around A-type stars is larger than that of solar-type stars$\footnotemark{}$$\footnotetext{\scriptsize An A-type star has a typical blowout grain size of a few microns, whereas a solar-type star has a blowout grain size of 1 $\mu$m or smaller.}$, grains in a single disk around an A-type star would likely have a more limited distribution of grain sizes, and thus would not be able to mimic a spatially separated two temperature component dust belt. In addition, three stars (HD 15407, HD 76582, and HD 113766) are early or mid F-type, and are also likely to have a limited grain size distribution. We expect that these seven stars likely have two spatially separated belts. 

For the remaining two systems (HD 35650 and HD 23514), we calculated a temperature ratio (R$_{T}$=T$_{d,2}$/T$_{d,1}$, see Table 3) to determine if the dust emission is coming from two separate belts. From Figure 4 of \citet{Kennedy_2014}, we see that single-belt systems with two temperature components tend to have small (R$_{T}$ $\lesssim$ 5) values of R$_{T}$. By contrast, spatially separated double-belt systems appear to have values of R$_{T}$ $\gtrsim$ 5; both HD 35650 (R$_{T}$=11.9) and HD 23514 (R$_{T}$=6.4) thus likely represents true spatially separated belts (see Figure \ref{fig:doubles}).

\section{Comparison with Previous Studies}\label{previous}
We compared our results to previous studies in the literature. In particular, we compare literature results with our SED-fitting results only, since none of the stars that follow have ever previously been resolved.

\subsection{HD 15407}
This F5/K2 binary system has an age of 80 Myr, derived from high resolution measurements of lithium in the photospheres of both components, X-ray data, and UVW space velocities \citep{Melis_2010}. Si emission features are seen in the IRS spectra \citep{Mittal_2015}. Melis et al. predicted (based on the IRS spectrum and the IRAS upper limits) that there would be no cold dust detected around HD 15407A. \citet{Olofsson_2012} performed a detailed analysis of the emission features in the IRS spectra, and found that HD 15407A should have an extended belt (from 0.4-19.2 AU) with a grain size distribution proportional to n$^{-3.1}$. This is shallower than the typical -3.5 value assumed for most debris disks. Olofsson et al. calculate a dust mass of $\sim$ 7.7E-05 M$_{\Earth}$. \citet{Fujiwara_2012} estimated that the dust should lie between 0.6 and 1.0 AU.

We find a double-belt system around HD 15407 (Figure 1). Both dust components are hot (T$\sim$334K, 1022K) and lie within 1 AU of the central star. These results are consistent with the prediction of a warm, extended belt \citep{Fujiwara_2012}. \citet{Olofsson_2012} studied the mid-IR emission features of HD 15407, and (based on the width of the disk), predicted that a cold dust belt should be found, that is responsible for feeding dust to the observed hot dust belt. The absence of a cold dust component in the Herschel observations implies that either there is not enough dust to be observable with Herschel, or the dust is unrealistically cold ($<$10K). The lack of a substantial dust belt would be consistent with a proposed model of transient dust.

\subsection{HD 23514}
HD 23514 is a Pleiades member with an IR excess detected by IRAS and Spitzer \citep{Rhee_2008}.  Rhee et al. report a dust temperature of 750 K and an exceptionally high fractional IR luminosity of 2.0E-02. Notably, they also report the results of observations with the Michelle Spectrograph on Gemini North, that show an unusual emission feature peaking around 9 $\mu$m. This feature (attributed to SiO$_{2}$) may indicate a recent major high-velocity collision involving a differentiated terrestrial planet \citep{Lisse_2009}. Rhee et al. found a lack of olivine and pyroxene in the spectra, which suggests that a steady-state collisional cascade involving an asteroid belt is likely not the cause of the observed dust. This model is only strengthened by the fact that we detected no cold dust component with Herschel, and is consistent with our calculations of $\tau$/$\tau$$_{max}$$\sim$300. We fit two dust components to the observed emission at 168K and 1082K. 

\citet{Rodriguez_2012b} report a substellar companion to HD23514 with a mass of $\sim$ 0.06M$_{\sun}$ at a separation of $\sim$360 AU. \citet{Zuckerman_2015} notes that several stars with dominant warm dust components were later discovered to be members of wide binary systems. The connection between the wide binary and the presence of warm dust is not well understood at this time.

\subsection{HD 35650}
This star is a known AB Dor member \citep{daSilva_2009}, with age $\sim$100Myr. It was observed by Spitzer, and was found to have a 70 $\mu$m excess only. \citet{Zuckerman_2011} report a dust temperature of 60 K and $\tau$ $\sim$1.7E-04. We find that the cold ($\sim$45 K) dust has $\tau$=1.5E-04, consistent with previous results. We also find that the mid-IR flux could be fit with an additional hot dust component at 536 K (Figure 1). HD 35650 was observed by the Near-Infrared Coronagraphic Imager (NICI) with angular differential imaging (ADI), but was not found to have a substellar companion \citep{Biller_2013}. 

\subsection{HD 43989}
HD 43989 is a member of Tuc Hor \citep{Zuckerman_2004}, and as such has an age of $\sim$40 Myr \citep{Kraus_2014}. \citet{Zuckerman_2011} report Spitzer observations that show an excess at 24 $\mu$m and in the IRS spectrum, but only an upper limit at 70 $\mu$m. We find a dust temperature of 112 K and a $\tau$ of 7.4E-05.  The star was observed for a planetary companion by the Multiple Mirror Telescope (MMT) with spectral differential imaging (SDI), but no such companion was found \citep{Biller_2007}. 

\subsection{HD 54341}
This A0 star was presented by \citet{Rhee_2007}. The system age is unreliable; the age based on isochrones disagrees with the age based on UVW space motions (while the isochrones suggest an age of 10 Myr, the UVWs suggest an older age). At the time, there was no MIPS data point, and the dust temperature and luminosity quoted in \cite{Rhee_2007} is unrealistic. It was also observed with adaptive optics at Lick Observatory and found not to be a member of a binary system \citep{Rodriguez_2012}. The system was observed by Spitzer, and \citet{Chen_2014} fit a two-component dust belt, with dust temperatures of 246K and 60K, and fractional IR luminosities of 1.9E-05 and 1.7E-04, respectively. While the $\chi$$^{2}$ values for a single- and double- belt fit were similar, we chose to fit a double-belt to be consistent with Chen et al.'s results. We find that the two belts have temperatures of 168 K and 60 K. The discrepancy in the warm dust temperatures is likely due to the fact that we often did not weight the IRAS photometry highly, due to the fact that IRAS data points are typically inconsistent with other, more reliable photometry. HD 54341 is one of our resolved systems; we find that the resolved disk size ($\sim$185 AU) is almost twice as big as the blackbody disk size ($\sim$106 AU). 

\subsection{HD 76543}
This A5 star was initially reported by \cite{Rhee_2007} to have an IRAS excess suggestive of a disk at 85 K with $\tau$=1.04E-04. Based on its location in UVW space (well outside of the ``good box" of \citealt{Zuckerman_2004}) and its location on the HR diagram, this star is estimated to have an age of $\sim$400 Myr. The star was also observed with Spitzer. \citet{Chen_2014} report a two-component disk with a cool belt at 81 K and a warm belt at 146 K. They find fractional IR luminosities of the two systems of 3E-05 and 1.7E-05 for the cool and warm belts, respectively. The addition of Herschel data helps to constrain the dust temperatures; we now find two belts of dust - one at 49 K and one at 105 K with fractional IR luminosities of 4.8E-05 and 3.6E-05, respectively. Any discrepancy between our dust temperatures and those of \citet{Chen_2014} are likely due to the high number of free parameters needed to fit a double belt system; a different cool dust temperature would correspond to a different warm dust temperature. The dust temperature from \citet{Rhee_2007} does not take into account any Spitzer or Herschel data. 

\subsection{HD 76582}
This F0 star was reported by \cite{Rhee_2007} to have an IRAS excess indicative of a disk at 85 K and $\tau$=2.22E-04. This relatively high fractional IR luminosity is surprising, given its 300 Myr age. The star was also observed by Spitzer \citep{Chen_2014}. These observations revealed a two-component disk, with an inner belt at 466 K (akin to a zodiacal dust cloud) and an outer belt at 78 K (akin to a Kuiper Belt). Our Herschel observations constrain the long wavelength behavior of the cold belt. We also obtain a two-component fit to our SEDs, with an inner belt at 132 K (possibly an asteroid belt analog) and an outer belt at 51 K (a close analog of the Kuiper Belt). The cold dust emission appears to peak at Herschel wavelengths; after a fit with a 51 K curve, dust at $\sim$450 K does not appear to be needed. Longer wavelength data will help to constrain the grain properties and dust mass. This star was targeted for sub-mm follow-up by the DUNES team \citep{Marshall_2016}.

\subsection{HD 84870}
 HD 84870 was first reported by \citet{Rhee_2007} to host a 100 Myr old debris disk with an IRAS excess indicative of a dust belt at 85 K. Rhee et al. listed this A3 star as a binary, but \citet{Mason_2013} describe the two stars in question as an optical pair (sep$\sim$30$\arcsec$) based on a study of their relative motions. \citet{Mittal_2015} report a slight Si emission at 10 $\mu$m (5$\sigma$) and 20 $\mu$m (12$\sigma$). They also report 70 $\mu$m photometry that suggests a double-belt architecture, with a hot inner belt at 553 K and a cool outer belt at 67 K. Our Herschel observations are slightly inconsistent with the Spitzer data point at 70 $\mu$m, and we do not find a need for a double-belt system. We fit a single belt at 54 K with $\tau$=4.7E-04.
 
\subsection{HD 85672}
This A0 star was first reported as a debris disk by \citet{Rhee_2007} with a cool belt at 79 K and $\tau$=4.82E-04. Our Herschel results (combined with new photometry from WISE) better constrain the temperature of the observed dust. We found that the long-wavelength photometry required a modified blackbody fit ($\lambda$$_{0}$=107, $\beta$=1.1), resulting in a best-fit dust temperature of 77 K and $\tau$=6.0E-04.

\subsection{HD 99945}
HD 99945 was initially reported by \citet{Rhee_2007} as a new debris disk discovered by IRAS. While our Herschel photometry at 100 $\mu$m is inconsistent with the IRAS data point at 60 $\mu$m, we are able to fit a belt of cold dust to the emission (T=38 K). Due to the low temperature, we examined this A2 star to determine if the emission might the due to contamination with a background galaxy or IR cirrus (see Sections 4.1 and 4.2). Due to the high far-IR flux and the fact that the emission is slightly too warm do be accounted for by a background galaxy at z$\sim$1, we believe that the excess emission indicates a belt of dust at 167 AU from the host star. 

\subsection{HD 113766}
HD 113766 is a well-known, extremely bright debris disk with a hot dust component. The mid-IR photometry has been well-covered in the literature. Si emission features are seen in the IRS spectra \citep{Mittal_2015}. \citet{Chen_2005} reported the Spitzer results, and found a single belt of dust at 330 K. \cite{Olofsson_2013} observed this star with VLTI/MIDI and Herschel/PACS, and determined that, in order to fit the mid-IR data simultaneously with the far-IR photometry, two spatially separated belts were needed: a hot belt within 1 AU, made of small grains which cause emission features in the mid-IR, and an outer belt at 9-13 AU. Our results independently confirm this two-belt structure, and are consistent with the findings of \cite{Olofsson_2013} regarding the transient nature of the dust. We report an inner belt at 0.2 AU and an outer belt at 2.3 AU. Our estimate for the semimajor axis of the cooler dust is smaller than that of \citet{Olofsson_2013}, though Oloffson et al. used a modified blackbody fit to the dust, and attempted to fit their VLTI/MIDI data simultaneously. As discussed in \cite{Olofsson_2013}, the binary companion to HD 113766A is likely too far from the disk to have any serious influence on the stirring of the disk material, though \citet{Zuckerman_2015} point out a possible relationship between distant companions and copious amounts of warm dust.

\subsection{HD 121191 and HD 131488}\label{HD121191}
These A-type stars were OT-1 targets. \citet{Melis_2013} report mid-IR imaging and spectroscopy from Gemini/T-ReCS. Most notably, they discovered that HD 121191 and HD 131488 both have unusual emission features that peak shortward of 8 $\mu$m. This emission feature does not appear to be due silica, olivines, or pyroxenes, all which show emission features peaking between 8-12 $\mu$m. Our Herschel observations confirm the two-belt structure of the debris disk around HD 121191, with a hot inner disk at 555 K and a warm disk at 118 K. There does not seem to be a cold dust component in this system. Melis et al. fit a two-belt debris disk to the SED of HD 131488, with one inner disk at 750 K and an outer disk at 100 K. Our Herschel observations confirm this double-belt nature of the debris disk, but with an inner belt at 570 K and an outer belt at 94 K. It should be noted that, while we believe HD 121191 is at least marginally resolved in the Herschel images, confirmation is needed to consider it to be a truly resolved disk since it appears as an extreme outlier in Figure \ref{fig:radii}.

\subsection{HD 124718}
This G5 star was initially reported by \citet{Rhee_2007} to have an IRAS excess indicative of a dust belt at 85 K and with a large fractional IR luminosity ($\tau$=2.11E-03), which was surprising given its old age ($>$500 Myr). WISE mid-IR photometry reveals an excess at 22 $\mu$m. Our Herschel observations are inconsistent with the IRAS data, and demonstrate that there is no accompanying cold dust belt. We fit a dust temperature of 179 K and $\tau$=2.1E-04. This discrepancy is unsurprising, since the dust temperature and luminosity reported by Rhee et al. were based on a single data point in excess, before WISE results were available.

\subsection{BD+20 307}
The excess IR emission around this well-studied close G-type binary star system was first reported by \citet{Song_2005} primarily based on IRAS photometry at 12 and 25 $\mu$m (the 60 and 100 $\mu$m IRAS observations only returned upper limits). \citet{Zuckerman_2008} obtained optical spectra to better determine the system age. Their results suggest that BD+20 307 could, in fact, be several Gyr old. Thus, the massive quantity of hot dust is not a result of ongoing planet formation, but rather could be the result of a collision between two mature planets in the terrestrial planet zone. The absence of cold dust was also noted in \citet{Weinberger_2011}. The Herschel SED confirms that there is no cold dust emitting at far-IR wavelengths, and helps to constrain the temperature of the dust (417 K) and to lend support to the planet-collision model. We did not attempt to fit the Spitzer IRS spectrum, since it is dominated by emission features \citep{Mittal_2015}. Thus, we adopt the $\tau$ value of 0.032 from \citet{Weinberger_2011}, since our estimate of $\tau$ ($\sim$0.01) does not include flux from the strong silicate emission feature.

\section{Conclusions}
We observed 24 stars with the PACS camera on the Herschel Space Observatory. Two infrared sources detected with Herschel are offset from the target coordinates by $>$2$\arcsec$. These are unlikely to be due to dusty debris belts. Three targets were not detected with Herschel, and are non-excess stars. One star was detected with Herschel, but shows no evidence of an IR excess. Two stars have low IR fluxes, and could possibly be explained by contamination by a background galaxy at z$\sim$1. One target star is clearly contaminated by a background galaxy. The remaining 15 stars were examined to determine dust properties and the possibility of ongoing planet formation. A summary of the results can be found in Table 5.

\begin{itemize}
\item Nine stars  (HD 15407, HD 23514, HD 35650, HD 54341, HD 76543, HD 76582, HD 113766, HD 121191, and HD 131488) appear to have dust components at two temperatures according to SED fitting. All appear to have spatially separated dust belts. 

\item Three stars (HD 15407, HD 113766, and BD+20 307)  have disks that cannot be explained by a steady-state collisional process alone. Two other stars (HD 23514 and HD 124718) are likely explained by a transient process. The other debris disks could be explained by steady-state collisional processes.

\item Dust belts at HD 54341, HD 84870, HD 85672, and HD 121191 are at large enough distances from their host stars that the dust could be stirred by a yet-unseen planet or binary companion. Stirring of these disks by putative 1000 km-sized bodies may be insufficient to explain the dust production in these systems.

\item The stars HD 54341, HD 76543, HD 76582, HD 84870, HD 85672, and HD 99945 are spatially resolved (or probably resolved as in the case of HD 121191), and a comparison between their blackbody radii (from SED-fitting) and resolved radii show that the latter are typically larger than the former, perhaps by up to a factor as large as 10 (HD 121191). 

\item Six stars (HD 15407, HD 23514, HD113766, HD 121191, HD 124718, and BD+20 307) show no evidence of cold ($<$100 K) dust in the Herschel data.

\end{itemize}

\section{Acknowledgements}
This research has made use of the Exoplanet Orbit Database, the Exoplanet Data Explorer at exoplanets.org, the SIMBAD database, and VIZIER search engine, operated by CDS in France. Partial support for this work was provided by a NASA grant to UCLA and by an NSF Graduate Research Fellowship to Laura Vican. We also thank the referee for their helpful and thorough feedback.

\bibliographystyle{apj}
\bibliography{Herschel}

\begin{thebibliography}{}
\expandafter\ifx\csname natexlab\endcsname\relax\def\natexlab#1{#1}\fi

\bibitem[{{Aumann} {et~al.}(1984){Aumann}, {Beichman}, {Gillett}, {de Jong},
  {Houck}, {Low}, {Neugebauer}, {Walker}, \& {Wesselius}}]{Aumann_1984}
{Aumann}, H.~H., {Beichman}, C.~A., {Gillett}, F.~C., {et~al.} 1984, \apjl,
  278, L23

\bibitem[{{Ballering} {et~al.}(2013){Ballering}, {Rieke}, {Su}, \&
  {Montiel}}]{Ballering_2013}
{Ballering}, N.~P., {Rieke}, G.~H., {Su}, K.~Y.~L., \& {Montiel}, E. 2013,
  \apj, 775, 55

\bibitem[{{Biller} {et~al.}(2007){Biller}, {Close}, {Masciadri}, {Nielsen},
  {Lenzen}, {Brandner}, {McCarthy}, {Hartung}, {Kellner}, {Mamajek}, {Henning},
  {Miller}, {Kenworthy}, \& {Kulesa}}]{Biller_2007}
{Biller}, B.~A., {Close}, L.~M., {Masciadri}, E., {et~al.} 2007, \apjs, 173,
  143

\bibitem[{{Biller} {et~al.}(2013){Biller}, {Liu}, {Wahhaj}, {Nielsen},
  {Hayward}, {Males}, {Skemer}, {Close}, {Chun}, {Ftaclas}, {Clarke}, {Thatte},
  {Shkolnik}, {Reid}, {Hartung}, {Boss}, {Lin}, {Alencar}, {de Gouveia Dal
  Pino}, {Gregorio-Hetem}, \& {Toomey}}]{Biller_2013}
{Biller}, B.~A., {Liu}, M.~C., {Wahhaj}, Z., {et~al.} 2013, \apj, 777, 160

\bibitem[{{Booth} {et~al.}(2013){Booth}, {Kennedy}, {Sibthorpe}, {Matthews},
  {Wyatt}, {Duch{\^e}ne}, {Kavelaars}, {Rodriguez}, {Greaves}, {Koning},
  {Vican}, {Rieke}, {Su}, {Moro-Mart{\'{\i}}n}, \& {Kalas}}]{Booth_2013}
{Booth}, M., {Kennedy}, G., {Sibthorpe}, B., {et~al.} 2013, \mnras, 428, 1263

\bibitem[{{Boulade} {et~al.}(1995){Boulade}, {Cesarsky}, {Cretolle}, {Rio},
  {Roy}, {Vigroux}, \& {Cesarsky}}]{Boulade_1995}
{Boulade}, O., {Cesarsky}, C.~J., {Cretolle}, J., {et~al.} 1995, in Society of
  Photo-Optical Instrumentation Engineers (SPIE) Conference Series, Vol. 2475,
  Infrared Detectors and Instrumentation for Astronomy, ed. A.~M. {Fowler},
  360--366

\bibitem[{{Bryden} {et~al.}(2009){Bryden}, {Beichman}, {Carpenter}, {Rieke},
  {Stapelfeldt}, {Werner}, {Tanner}, {Lawler}, {Wyatt}, {Trilling}, {Su},
  {Blaylock}, \& {Stansberry}}]{Bryden_2009}
{Bryden}, G., {Beichman}, C.~A., {Carpenter}, J.~M., {et~al.} 2009, \apj, 705,
  1226

\bibitem[{{Casagrande} {et~al.}(2011){Casagrande}, {Sch{\"o}nrich}, {Asplund},
  {Cassisi}, {Ram{\'{\i}}rez}, {Mel{\'e}ndez}, {Bensby}, \&
  {Feltzing}}]{Casagrande_2011}
{Casagrande}, L., {Sch{\"o}nrich}, R., {Asplund}, M., {et~al.} 2011, \aap, 530,
  A138

\bibitem[{{Chen} \& {Jura}(2001)}]{Chen_2001}
{Chen}, C.~H., \& {Jura}, M. 2001, \apjl, 560, L171

\bibitem[{{Chen} {et~al.}(2005){Chen}, {Jura}, {Gordon}, \&
  {Blaylock}}]{Chen_2005}
{Chen}, C.~H., {Jura}, M., {Gordon}, K.~D., \& {Blaylock}, M. 2005, \apj, 623,
  493

\bibitem[{{Chen} {et~al.}(2014){Chen}, {Mittal}, {Kuchner}, {Forrest}, {Lisse},
  {Manoj}, {Sargent}, \& {Watson}}]{Chen_2014}
{Chen}, C.~H., {Mittal}, T., {Kuchner}, M., {et~al.} 2014, \apjs, 211, 25

\bibitem[{{Chen} {et~al.}(2009){Chen}, {Sheehan}, {Watson}, {Manoj}, \&
  {Najita}}]{Chen_2009}
{Chen}, C.~H., {Sheehan}, P., {Watson}, D.~M., {Manoj}, P., \& {Najita}, J.~R.
  2009, \apj, 701, 1367

\bibitem[{{Cutri} {et~al.}(2012){Cutri}, {Wright}, {Conrow}, {Bauer},
  {Benford}, {Brandenburg}, {Dailey}, {Eisenhardt}, {Evans}, {Fajardo-Acosta},
  {Fowler}, {Gelino}, {Grillmair}, {Harbut}, {Hoffman}, {Jarrett},
  {Kirkpatrick}, {Leisawitz}, {Liu}, {Mainzer}, {Marsh}, {Masci}, {McCallon},
  {Padgett}, {Ressler}, {Royer}, {Skrutskie}, {Stanford}, {Wyatt}, {Tholen},
  {Tsai}, {Wachter}, {Wheelock}, {Yan}, {Alles}, {Beck}, {Grav}, {Masiero},
  {McCollum}, {McGehee}, {Papin}, \& {Wittman}}]{Cutri_2012}
{Cutri}, R.~M., {Wright}, E.~L., {Conrow}, T., {et~al.} 2012, {Explanatory
  Supplement to the WISE All-Sky Data Release Products}, Tech. rep.

\bibitem[{{da Silva} {et~al.}(2009){da Silva}, {Torres}, {de La Reza}, {Quast},
  {Melo}, \& {Sterzik}}]{daSilva_2009}
{da Silva}, L., {Torres}, C.~A.~O., {de La Reza}, R., {et~al.} 2009, \aap, 508,
  833

\bibitem[{{de Graauw} {et~al.}(2008){de Graauw}, {Whyborn}, {Helmich},
  {Dieleman}, {Roelfsema}, {Caux}, {Phillips}, {Stutzki}, {Beintema}, {Benz},
  {Biver}, {Boogert}, {Boulanger}, {Cherednichenko}, {Coeur-Joly}, {Comito},
  {Dartois}, {de Jonge}, {de Lange}, {Delorme}, {DiGiorgio}, {Dubbeldam},
  {Edwards}, {Fich}, {G{\"u}sten}, {Herpin}, {Honingh}, {Huisman}, {Jacobs},
  {Jellema}, {Kawamura}, {Kester}, {Klapwijk}, {Klein}, {Kooi}, {Krieg},
  {Kramer}, {Kruizenga}, {Laauwen}, {Larsson}, {Leinz}, {Liseau}, {Lord},
  {Luinge}, {Marston}, {Merkel}, {Moreno}, {Morris}, {Murphy}, {Naber},
  {Planesas}, {Martin-Pintado}, {Olberg}, {Orleanski}, {Ossenkopf}, {Pearson},
  {Perault}, {Phillip}, {Rataj}, {Ravera}, {Saraceno}, {Schieder},
  {Schmuelling}, {Szczerba}, {Shipman}, {Teyssier}, {Vastel}, {Visser},
  {Wildeman}, {Wafelbakker}, {Ward}, {Higgins}, {Aarts}, {Tielens}, \&
  {Zaal}}]{deGraauw_2008}
{de Graauw}, T., {Whyborn}, N., {Helmich}, F., {et~al.} 2008, in Society of
  Photo-Optical Instrumentation Engineers (SPIE) Conference Series, Vol. 7010,
  Society of Photo-Optical Instrumentation Engineers (SPIE) Conference Series,
  4

\bibitem[{{Dominik} \& {Decin}(2003)}]{Dominik_2003}
{Dominik}, C., \& {Decin}, G. 2003, \apj, 598, 626

\bibitem[{{Fujiwara} {et~al.}(2012){Fujiwara}, {Onaka}, {Yamashita},
  {Ishihara}, {Kataza}, {Fukagawa}, {Takeda}, \& {Murakami}}]{Fujiwara_2012}
{Fujiwara}, H., {Onaka}, T., {Yamashita}, T., {et~al.} 2012, \apjl, 749, L29

\bibitem[{{G{\'a}sp{\'a}r} \& {Rieke}(2014)}]{Gaspar_2014}
{G{\'a}sp{\'a}r}, A., \& {Rieke}, G.~H. 2014, \apj, 784, 33

\bibitem[{{Griffin} {et~al.}(2008){Griffin}, {Swinyard}, {Vigroux}, {Abergel},
  {Ade}, {Andr{\'e}}, {Baluteau}, {Bock}, {Franceschini}, {Gear}, {Glenn},
  {Huang}, {Griffin}, {King}, {Lellouch}, {Naylor}, {Oliver}, {Olofsson},
  {Perez-Fournon}, {Page}, {Rowan-Robinson}, {Saraceno}, {Sawyer}, {Wright},
  {Zavagno}, {Abreu}, {Bendo}, {Dowell}, {Dowell}, {Ferlet}, {Fulton},
  {Hargrave}, {Laurent}, {Leeks}, {Lim}, {Lu}, {Nguyen}, {Pearce},
  {Polehampton}, {Rizzo}, {Schulz}, {Sidher}, {Smith}, {Spencer}, {Valtchanov},
  {Woodcraft}, {Xu}, \& {Zhang}}]{Griffin_2008}
{Griffin}, M., {Swinyard}, B., {Vigroux}, L., {et~al.} 2008, in Society of
  Photo-Optical Instrumentation Engineers (SPIE) Conference Series, Vol. 7010,
  Society of Photo-Optical Instrumentation Engineers (SPIE) Conference Series,
  6

\bibitem[{{Han} {et~al.}(2014){Han}, {Wang}, {Wright}, {Feng}, {Zhao},
  {Fakhouri}, {Brown}, \& {Hancock}}]{exoplanets}
{Han}, E., {Wang}, S.~X., {Wright}, J.~T., {et~al.} 2014, \pasp, 126, 827

\bibitem[{{Hartmann} \& {Davis}(1975)}]{Hartmann_1975}
{Hartmann}, W.~K., \& {Davis}, D.~R. 1975, Icarus, 24, 504

\bibitem[{{Hauschildt} {et~al.}(1999){Hauschildt}, {Allard}, \&
  {Baron}}]{Hauschildt_1999}
{Hauschildt}, P.~H., {Allard}, F., \& {Baron}, E. 1999, \apj, 512, 377

\bibitem[{{Houck} {et~al.}(2004){Houck}, {Roellig}, {Van Cleve}, {Forrest},
  {Herter}, {Lawrence}, {Matthews}, {Reitsema}, {Soifer}, {Watson}, {Weedman},
  {Huisjen}, {Troeltzsch}, {Barry}, {Bernard-Salas}, {Blacken}, {Brandl},
  {Charmandaris}, {Devost}, {Gull}, {Hall}, {Henderson}, {Higdon}, {Pirger},
  {Schoenwald}, {Sloan}, {Uchida}, {Appleton}, {Armus}, {Burgdorf},
  {Fajardo-Acosta}, {Grillmair}, {Ingalls}, {Morris}, \&
  {Teplitz}}]{Houck_2004}
{Houck}, J.~R., {Roellig}, T.~L., {Van Cleve}, J., {et~al.} 2004, in Society of
  Photo-Optical Instrumentation Engineers (SPIE) Conference Series, Vol. 5487,
  Optical, Infrared, and Millimeter Space Telescopes, ed. J.~C. {Mather},
  62--76

\bibitem[{{Jackson} \& {Wyatt}(2012)}]{Jackson_2012}
{Jackson}, A.~P., \& {Wyatt}, M.~C. 2012, \mnras, 425, 657

\bibitem[{{Kennedy} \& {Wyatt}(2014)}]{Kennedy_2014}
{Kennedy}, G.~M., \& {Wyatt}, M.~C. 2014, \mnras, 444, 3164

\bibitem[{{Kenyon} \& {Bromley}(2005)}]{Kenyon_2005}
{Kenyon}, S.~J., \& {Bromley}, B.~C. 2005, \aj, 130, 269

\bibitem[{{Kenyon} \& {Bromley}(2008)}]{Kenyon_2008}
---. 2008, \apjs, 179, 451

\bibitem[{{K{\'o}sp{\'a}l} {et~al.}(2013){K{\'o}sp{\'a}l}, {Mo{\'o}r},
  {Juh{\'a}sz}, {{\'A}brah{\'a}m}, {Apai}, {Csengeri}, {Grady}, {Henning},
  {Hughes}, {Kiss}, {Pascucci}, \& {Schmalzl}}]{Kospal_2013}
{K{\'o}sp{\'a}l}, {\'A}., {Mo{\'o}r}, A., {Juh{\'a}sz}, A., {et~al.} 2013,
  \apj, 776, 77

\bibitem[{{Kraus} {et~al.}(2014){Kraus}, {Shkolnik}, {Allers}, \&
  {Liu}}]{Kraus_2014}
{Kraus}, A.~L., {Shkolnik}, E.~L., {Allers}, K.~N., \& {Liu}, M.~C. 2014, \aj,
  147, 146

\bibitem[{{Lisse} {et~al.}(2008){Lisse}, {Chen}, {Wyatt}, \&
  {Morlok}}]{Lisse_2008}
{Lisse}, C.~M., {Chen}, C.~H., {Wyatt}, M.~C., \& {Morlok}, A. 2008, \apj, 673,
  1106

\bibitem[{{Lisse} {et~al.}(2009){Lisse}, {Chen}, {Wyatt}, {Morlok}, {Song},
  {Bryden}, \& {Sheehan}}]{Lisse_2009}
{Lisse}, C.~M., {Chen}, C.~H., {Wyatt}, M.~C., {et~al.} 2009, \apj, 701, 2019

\bibitem[{{Magnelli} {et~al.}(2013){Magnelli}, {Popesso}, {Berta}, {Pozzi},
  {Elbaz}, {Lutz}, {Dickinson}, {Altieri}, {Andreani}, {Aussel},
  {B{\'e}thermin}, {Bongiovanni}, {Cepa}, {Charmandaris}, {Chary}, {Cimatti},
  {Daddi}, {F{\"o}rster Schreiber}, {Genzel}, {Gruppioni}, {Harwit}, {Hwang},
  {Ivison}, {Magdis}, {Maiolino}, {Murphy}, {Nordon}, {Pannella}, {P{\'e}rez
  Garc{\'{\i}}a}, {Poglitsch}, {Rosario}, {Sanchez-Portal}, {Santini}, {Scott},
  {Sturm}, {Tacconi}, \& {Valtchanov}}]{Magnelli_2013}
{Magnelli}, B., {Popesso}, P., {Berta}, S., {et~al.} 2013, \aap, 553, A132

\bibitem[{{Marois} {et~al.}(2008){Marois}, {Macintosh}, {Barman}, {Zuckerman},
  {Song}, {Patience}, {Lafreni{\`e}re}, \& {Doyon}}]{Marois_2008}
{Marois}, C., {Macintosh}, B., {Barman}, T., {et~al.} 2008, Science, 322, 1348

\bibitem[{{Marois} {et~al.}(2010){Marois}, {Zuckerman}, {Konopacky},
  {Macintosh}, \& {Barman}}]{Marois_2010}
{Marois}, C., {Zuckerman}, B., {Konopacky}, Q.~M., {Macintosh}, B., \&
  {Barman}, T. 2010, \nat, 468, 1080

\bibitem[{{Marshall} {et~al.}(2016){Marshall}, {Booth}, {Holland}, {Matthews},
  {Greaves}, \& {Zuckerman}}]{Marshall_2016}
{Marshall}, J.~P., {Booth}, M., {Holland}, W., {et~al.} 2016, \mnras, 459, 2893

\bibitem[{{Mason} {et~al.}(2013){Mason}, {Wycoff}, {Hartkopf}, {Douglass}, \&
  {Worley}}]{Mason_2013}
{Mason}, B.~D., {Wycoff}, G.~L., {Hartkopf}, W.~I., {Douglass}, G.~G., \&
  {Worley}, C.~E. 2013, VizieR Online Data Catalog, 1, 2026

\bibitem[{{Matthews} {et~al.}(2010){Matthews}, {Sibthorpe}, {Kennedy},
  {Phillips}, {Churcher}, {Duch{\^e}ne}, {Greaves}, {Lestrade}, {Moro-Martin},
  {Wyatt}, {Bastien}, {Biggs}, {Bouvier}, {Butner}, {Dent}, {di Francesco},
  {Eisl{\"o}ffel}, {Graham}, {Harvey}, {Hauschildt}, {Holland}, {Horner},
  {Ibar}, {Ivison}, {Johnstone}, {Kalas}, {Kavelaars}, {Rodriguez}, {Udry},
  {van der Werf}, {Wilner}, \& {Zuckerman}}]{Matthews_2010}
{Matthews}, B.~C., {Sibthorpe}, B., {Kennedy}, G., {et~al.} 2010, \aap, 518,
  L135

\bibitem[{{McDonald} {et~al.}(2012){McDonald}, {Zijlstra}, \&
  {Boyer}}]{McDonald_2012}
{McDonald}, I., {Zijlstra}, A.~A., \& {Boyer}, M.~L. 2012, \mnras, 427, 343

\bibitem[{{Melis} {et~al.}(2010){Melis}, {Zuckerman}, {Rhee}, \&
  {Song}}]{Melis_2010}
{Melis}, C., {Zuckerman}, B., {Rhee}, J.~H., \& {Song}, I. 2010, \apjl, 717,
  L57

\bibitem[{{Melis} {et~al.}(2013){Melis}, {Zuckerman}, {Rhee}, {Song}, {Murphy},
  \& {Bessell}}]{Melis_2013}
{Melis}, C., {Zuckerman}, B., {Rhee}, J.~H., {et~al.} 2013, \apj, 778, 12

\bibitem[{{Mittal} {et~al.}(2015){Mittal}, {Chen}, {Jang-Condell}, {Manoj},
  {Sargent}, {Watson}, \& {Lisse}}]{Mittal_2015}
{Mittal}, T., {Chen}, C.~H., {Jang-Condell}, H., {et~al.} 2015, \apj, 798, 87

\bibitem[{{Mo{\'o}r} {et~al.}(2015){Mo{\'o}r}, {K{\'o}sp{\'a}l},
  {{\'A}brah{\'a}m}, {Apai}, {Balog}, {Grady}, {Henning}, {Juh{\'a}sz}, {Kiss},
  {Krivov}, {Pawellek}, \& {Szab{\'o}}}]{Moor_2015}
{Mo{\'o}r}, A., {K{\'o}sp{\'a}l}, {\'A}., {{\'A}brah{\'a}m}, P., {et~al.} 2015,
  \mnras, 447, 577

\bibitem[{{Morales} {et~al.}(2013){Morales}, {Bryden}, {Werner}, \&
  {Stapelfeldt}}]{Morales_2013}
{Morales}, F.~Y., {Bryden}, G., {Werner}, M.~W., \& {Stapelfeldt}, K.~R. 2013,
  \apj, 776, 111

\bibitem[{{Morales} {et~al.}(2011){Morales}, {Rieke}, {Werner}, {Bryden},
  {Stapelfeldt}, \& {Su}}]{Morales_2011}
{Morales}, F.~Y., {Rieke}, G.~H., {Werner}, M.~W., {et~al.} 2011, \apjl, 730,
  L29

\bibitem[{{Moro-Mart{\'{\i}}n} {et~al.}(2015){Moro-Mart{\'{\i}}n}, {Marshall},
  {Kennedy}, {Sibthorpe}, {Matthews}, {Eiroa}, {Wyatt}, {Lestrade},
  {Maldonado}, {Rodriguez}, {Greaves}, {Montesinos}, {Mora}, {Booth},
  {Duch{\^e}ne}, {Wilner}, \& {Horner}}]{Moro_2015}
{Moro-Mart{\'{\i}}n}, A., {Marshall}, J.~P., {Kennedy}, G., {et~al.} 2015,
  \apj, 801, 143

\bibitem[{{Nesvold} {et~al.}(2016){Nesvold}, {Naoz}, {Vican}, \&
  {Farr}}]{Nesvold_2016}
{Nesvold}, E.~R., {Naoz}, S., {Vican}, L., \& {Farr}, W.~M. 2016, ArXiv
  e-prints, arXiv:1603.08005

\bibitem[{{Olofsson} {et~al.}(2013){Olofsson}, {Henning}, {Nielbock},
  {Augereau}, {Juh{\`a}sz}, {Oliveira}, {Absil}, \& {Tamanai}}]{Olofsson_2013}
{Olofsson}, J., {Henning}, T., {Nielbock}, M., {et~al.} 2013, \aap, 551, A134

\bibitem[{{Olofsson} {et~al.}(2012){Olofsson}, {Juh{\'a}sz}, {Henning},
  {Mutschke}, {Tamanai}, {Mo{\'o}r}, \& {{\'A}brah{\'a}m}}]{Olofsson_2012}
{Olofsson}, J., {Juh{\'a}sz}, A., {Henning}, T., {et~al.} 2012, \aap, 542, A90

\bibitem[{{Ott}(2010)}]{2010ASPC..434..139O}
{Ott}, S. 2010, in Astronomical Society of the Pacific Conference Series, Vol.
  434, Astronomical Data Analysis Software and Systems XIX, ed. Y.~{Mizumoto},
  K.-I. {Morita}, \& M.~{Ohishi}, 139

\bibitem[{{Pawellek} {et~al.}(2014){Pawellek}, {Krivov}, {Marshall},
  {Montesinos}, {{\'A}brah{\'a}m}, {Mo{\'o}r}, {Bryden}, \&
  {Eiroa}}]{Pawellek_2014}
{Pawellek}, N., {Krivov}, A.~V., {Marshall}, J.~P., {et~al.} 2014, \apj, 792,
  65

\bibitem[{{Pilbratt} {et~al.}(2010){Pilbratt}, {Riedinger}, {Passvogel},
  {Crone}, {Doyle}, {Gageur}, {Heras}, {Jewell}, {Metcalfe}, {Ott}, \&
  {Schmidt}}]{Pilbratt_2010}
{Pilbratt}, G.~L., {Riedinger}, J.~R., {Passvogel}, T., {et~al.} 2010, \aap,
  518, L1

\bibitem[{{Plavchan} {et~al.}(2009){Plavchan}, {Werner}, {Chen}, {Stapelfeldt},
  {Su}, {Stauffer}, \& {Song}}]{Plavchan_2009}
{Plavchan}, P., {Werner}, M.~W., {Chen}, C.~H., {et~al.} 2009, \apj, 698, 1068

\bibitem[{{Poglitsch} {et~al.}(2008){Poglitsch}, {Waelkens}, {Bauer}, {Cepa},
  {Feuchtgruber}, {Henning}, {van Hoof}, {Kerschbaum}, {Krause}, {Renotte},
  {Rodriguez}, {Saraceno}, \& {Vandenbussche}}]{Poglitsch_2008}
{Poglitsch}, A., {Waelkens}, C., {Bauer}, O.~H., {et~al.} 2008, in Society of
  Photo-Optical Instrumentation Engineers (SPIE) Conference Series, Vol. 7010,
  Society of Photo-Optical Instrumentation Engineers (SPIE) Conference Series,
  5

\bibitem[{{Rameau} {et~al.}(2013){Rameau}, {Chauvin}, {Lagrange}, {Meshkat},
  {Boccaletti}, {Quanz}, {Currie}, {Mawet}, {Girard}, {Bonnefoy}, \&
  {Kenworthy}}]{Rameau_2013}
{Rameau}, J., {Chauvin}, G., {Lagrange}, A.-M., {et~al.} 2013, \apjl, 779, L26

\bibitem[{{Rhee} {et~al.}(2008){Rhee}, {Song}, \& {Zuckerman}}]{Rhee_2008}
{Rhee}, J.~H., {Song}, I., \& {Zuckerman}, B. 2008, \apj, 675, 777

\bibitem[{{Rhee} {et~al.}(2007){Rhee}, {Song}, {Zuckerman}, \&
  {McElwain}}]{Rhee_2007}
{Rhee}, J.~H., {Song}, I., {Zuckerman}, B., \& {McElwain}, M. 2007, \apj, 660,
  1556

\bibitem[{{Rieke} {et~al.}(2004){Rieke}, {Young}, {Engelbracht}, {Kelly},
  {Low}, {Haller}, {Beeman}, {Gordon}, {Stansberry}, {Misselt}, {Cadien},
  {Morrison}, {Rivlis}, {Latter}, {Noriega-Crespo}, {Padgett}, {Stapelfeldt},
  {Hines}, {Egami}, {Muzerolle}, {Alonso-Herrero}, {Blaylock}, {Dole}, {Hinz},
  {Le Floc'h}, {Papovich}, {P{\'e}rez-Gonz{\'a}lez}, {Smith}, {Su}, {Bennett},
  {Frayer}, {Henderson}, {Lu}, {Masci}, {Pesenson}, {Rebull}, {Rho}, {Keene},
  {Stolovy}, {Wachter}, {Wheaton}, {Werner}, \& {Richards}}]{Rieke_2004}
{Rieke}, G.~H., {Young}, E.~T., {Engelbracht}, C.~W., {et~al.} 2004, \apjs,
  154, 25

\bibitem[{{Rodriguez} {et~al.}(2015){Rodriguez}, {Duch{\^e}ne}, {Tom},
  {Kennedy}, {Matthews}, {Greaves}, \& {Butner}}]{Rodriguez_2015}
{Rodriguez}, D.~R., {Duch{\^e}ne}, G., {Tom}, H., {et~al.} 2015, \mnras, 449,
  3160

\bibitem[{{Rodriguez} {et~al.}(2012){Rodriguez}, {Marois}, {Zuckerman},
  {Macintosh}, \& {Melis}}]{Rodriguez_2012b}
{Rodriguez}, D.~R., {Marois}, C., {Zuckerman}, B., {Macintosh}, B., \& {Melis},
  C. 2012, \apj, 748, 30

\bibitem[{{Rodriguez} \& {Zuckerman}(2012)}]{Rodriguez_2012}
{Rodriguez}, D.~R., \& {Zuckerman}, B. 2012, \apj, 745, 147

\bibitem[{{Roy} {et~al.}(2010){Roy}, {Ade}, {Bock}, {Chapin}, {Devlin},
  {Dicker}, {Griffin}, {Gundersen}, {Halpern}, {Hargrave}, {Hughes}, {Klein},
  {Marsden}, {Martin}, {Mauskopf}, {Miville-Desch{\^e}nes}, {Netterfield},
  {Olmi}, {Patanchon}, {Rex}, {Scott}, {Semisch}, {Truch}, {Tucker}, {Tucker},
  {Viero}, \& {Wiebe}}]{Roy_2010}
{Roy}, A., {Ade}, P.~A.~R., {Bock}, J.~J., {et~al.} 2010, \apj, 708, 1611

\bibitem[{{Song} {et~al.}(2003){Song}, {Zuckerman}, \& {Bessell}}]{Song_2003}
{Song}, I., {Zuckerman}, B., \& {Bessell}, M.~S. 2003, \apj, 599, 342

\bibitem[{{Song} {et~al.}(2005){Song}, {Zuckerman}, {Weinberger}, \&
  {Becklin}}]{Song_2005}
{Song}, I., {Zuckerman}, B., {Weinberger}, A.~J., \& {Becklin}, E.~E. 2005,
  \nat, 436, 363

\bibitem[{{Su} {et~al.}(2015){Su}, {Morrison}, {Malhotra}, {Smith}, {Balog}, \&
  {Rieke}}]{Su_2015}
{Su}, K.~Y.~L., {Morrison}, S., {Malhotra}, R., {et~al.} 2015, \apj, 799, 146

\bibitem[{{Tetzlaff} {et~al.}(2011){Tetzlaff}, {Neuh{\"a}user}, \&
  {Hohle}}]{Tetzlaff_2011}
{Tetzlaff}, N., {Neuh{\"a}user}, R., \& {Hohle}, M.~M. 2011, \mnras, 410, 190

\bibitem[{{Weinberger} {et~al.}(2011){Weinberger}, {Becklin}, {Song}, \&
  {Zuckerman}}]{Weinberger_2011}
{Weinberger}, A.~J., {Becklin}, E.~E., {Song}, I., \& {Zuckerman}, B. 2011,
  \apj, 726, 72

\bibitem[{{Werner} {et~al.}(2004){Werner}, {Roellig}, {Low}, {Rieke}, {Rieke},
  {Hoffmann}, {Young}, {Houck}, {Brandl}, {Fazio}, {Hora}, {Gehrz}, {Helou},
  {Soifer}, {Stauffer}, {Keene}, {Eisenhardt}, {Gallagher}, {Gautier}, {Irace},
  {Lawrence}, {Simmons}, {Van Cleve}, {Jura}, {Wright}, \&
  {Cruikshank}}]{Werner_2004}
{Werner}, M.~W., {Roellig}, T.~L., {Low}, F.~J., {et~al.} 2004, \apjs, 154, 1

\bibitem[{{Williams} \& {Cieza}(2011)}]{Williams_2011}
{Williams}, J.~P., \& {Cieza}, L.~A. 2011, \araa, 49, 67

\bibitem[{{Wright} {et~al.}(2010){Wright}, {Eisenhardt}, {Mainzer}, {Ressler},
  {Cutri}, {Jarrett}, {Kirkpatrick}, {Padgett}, {McMillan}, {Skrutskie},
  {Stanford}, {Cohen}, {Walker}, {Mather}, {Leisawitz}, {Gautier}, {McLean},
  {Benford}, {Lonsdale}, {Blain}, {Mendez}, {Irace}, {Duval}, {Liu}, {Royer},
  {Heinrichsen}, {Howard}, {Shannon}, {Kendall}, {Walsh}, {Larsen}, {Cardon},
  {Schick}, {Schwalm}, {Abid}, {Fabinsky}, {Naes}, \& {Tsai}}]{Wright_2010}
{Wright}, E.~L., {Eisenhardt}, P.~R.~M., {Mainzer}, A.~K., {et~al.} 2010, \aj,
  140, 1868

\bibitem[{{Wyatt}(2008)}]{Wyatt_2008}
{Wyatt}, M.~C. 2008, \araa, 46, 339

\bibitem[{{Wyatt} {et~al.}(2007){Wyatt}, {Smith}, {Greaves}, {Beichman},
  {Bryden}, \& {Lisse}}]{Wyatt_2007}
{Wyatt}, M.~C., {Smith}, R., {Greaves}, J.~S., {et~al.} 2007, \apj, 658, 569

\bibitem[{{Zorec} \& {Royer}(2012)}]{Zorec_2012}
{Zorec}, J., \& {Royer}, F. 2012, \aap, 537, A120

\bibitem[{{Zuckerman}(2015)}]{Zuckerman_2015}
{Zuckerman}, B. 2015, \apj, 798, 86

\bibitem[{{Zuckerman} {et~al.}(2008){Zuckerman}, {Fekel}, {Williamson},
  {Henry}, \& {Muno}}]{Zuckerman_2008}
{Zuckerman}, B., {Fekel}, F.~C., {Williamson}, M.~H., {Henry}, G.~W., \&
  {Muno}, M.~P. 2008, \apj, 688, 1345

\bibitem[{{Zuckerman} {et~al.}(2012){Zuckerman}, {Melis}, {Rhee}, {Schneider},
  \& {Song}}]{Zuckerman_2012b}
{Zuckerman}, B., {Melis}, C., {Rhee}, J.~H., {Schneider}, A., \& {Song}, I.
  2012, \apj, 752, 58

\bibitem[{{Zuckerman} {et~al.}(2011){Zuckerman}, {Rhee}, {Song}, \&
  {Bessell}}]{Zuckerman_2011}
{Zuckerman}, B., {Rhee}, J.~H., {Song}, I., \& {Bessell}, M.~S. 2011, \apj,
  732, 61

\bibitem[{{Zuckerman} \& {Song}(2004)}]{Zuckerman_2004}
{Zuckerman}, B., \& {Song}, I. 2004, \apj, 603, 738

\bibitem[{{Zuckerman} \& {Song}(2012)}]{Zuckerman_2012}
---. 2012, \apj, 758, 77

\end{thebibliography}

%TABLE 1 -
\begin{deluxetable}{cccccccccccc}
\tablewidth{0pt}
\tablecaption{Stellar Parameters\label{table:stellar}}
\tabletypesize{\tiny}
\tablehead{
\colhead{HD}&
\colhead{HIP}&
\colhead{SpT}&
\colhead{d}&
\colhead{M$_{*}$$^{1}$}&
\colhead{R$_{*}$}&
\colhead{T$_{*}$}&
\colhead{L$_{*}$}&
\colhead{Age}&
\colhead{Age}&
\colhead{Age}&
\colhead{binary?}\\
\colhead{}&
\colhead{}&
\colhead{}&
\colhead{[pc]}&
\colhead{[M$_{\sun}$]}&
\colhead{[R$_{\sun}$]}&
\colhead{[K]}&
\colhead{[L$_{\sun}$]}&
\colhead{[Myr]}&
\colhead{Method$^{2}$}&
\colhead{Ref$^{3}$}&
\colhead{}}
\startdata
\multicolumn{12}{c}{Herschel Detections}\\
\hline
15407&11696& F5 &55&1.4&1.6&6500&4.0&80& a & M10& Y \\
23514&& F6 &135&1.4&1.08&6300&1.6&100& b & R08 &Y \\
35650&25283& K6 &18&0.8&0.63&4300&0.1&70&b&Z11& \\
43989&30030& G0 &49&1.1&1.05&6100&1.3&30& b & Z11 & \\
54341&34276& A0 &93&2.4&1.83&9500&24.1&10&c&R07 &\\
76543&43970& A5 &49&1.9&1.83&8200&13.4&400&c&R07& \\
76582&44001& F0 &49&1.7&1.68&7700&8.8&300&c&R07& \\
84870&48164& A3 &90&1.6&1.63&7600&7.8&100&c&R07&\\
85672&48541& A0 &93&1.7&1.51&8000&8.3&30&c&R07& \\
99945&56253& A2 &60&1.8&1.79&7600&9.4&300&c&R07& \\
113766&& F3 &120&1.9&1.9&6100&4.4&15&b&C05& Y \\
121191&& A5 &130&1.8&1.83&7700&10.4&10&b&M13& N \\
124718&69682& G5 &61&1&0.96&5900&0.99& $>$500 &a&S03& \\
131488&& A1 &150&2.2&2.06&8700&21.4&10&b&M13& N \\
BD+20 307 &8920& G0 &92&1.9&1.26&6100&1.9&1000&a&Z08& Y \\
\hline
\multicolumn{12}{c}{Herschel Non-Detections}\\
\hline
60234&36906& G0 &108&1.6&3.53$^{4}$&5900&13.3&600&d&R07& \\
70298&40938& F2 &71&&1.73&6500&4.7&$>$3000&d&R07& \\
72660&42028& A1 &100&&2.24&9500&36.1&200&c&R07& \\
132950&73512& K2 &30&&0.73&4800&0.25&3000&...&R07& \\
203562&105570& A3 &110&2.7&3.73$^{4}$&8800&6.5&600&c&R07&\\
\hline
\multicolumn{12}{c}{Contaminated Fields}\\
\hline
8558&6485& G6 &50&0.9&0.94&5800&0.9&30& b & Z11 &\\
13183&9892& G7 &50&0.9&0.95&5700&0.8&30& b & Z11 &\\
80425&45758& A5 &98&&2.54$^{4}$&7500&18.0&300&c&R07& \\
191692&99473& B9 &88&3.34&6.40&10500&2680&500&c&R07&Y \\
\enddata
\tablecomments{\tiny$^{1}$Stellar masses were taken from \citet{Chen_2014}, \citet{Casagrande_2011}, \citet{Zorec_2012}, or \citet{Tetzlaff_2011}.).
$^{2}$ Age methods: (a) Li abundance, (b) Association membership, (c) Location on an HR diagram, (d) Activity (X-ray luminosity). Note that these indicate the primary method used to identify stellar age, but supplementary methods may have been applied.$^{3}$ Refs: M10 = \citet{Melis_2010}, M13=\citet{Melis_2013}, R08=\citet{Rhee_2008},  Z11= \citet{Zuckerman_2011}, R07=\citet{Rhee_2007}, S03=\citet{Song_2003}, C05=\citet{Chen_2005}, Z08=\citet{Zuckerman_2008} $^{4}$These stars are likely not main-sequence, based on their radii.}
\end{deluxetable}

%TABLE 2 - Observations
\begin{deluxetable}{ccccc}
\tablewidth{0pt}
\tablecaption{Herschel Observations \label{observations}}
\tabletypesize{\tiny}
\tablehead{
\colhead{HD}&
\colhead{OT}&
\colhead{$\lambda$}&
\colhead{F [mJy]}&
\colhead{$\sigma$ [mJy]}}
\startdata
15407 & OT1bzuckerm1 & 70 & 55.7 & 4.5 \\
&& 100 & 24.4 & 3.8 \\
&& 160 & 52.6 & 14.9  \\
23514 & OT1bzuckerm1 & 70 & 24.8 & 3.5 \\
&& 100 & 83.5 & 3.2 \\
&& 160 & 60.7 & 15.2 \\
35650 & OT2bzuckerm2 &100 & 26.8 & 1.9 \\
&& 160 & 57.0 & 25.4  \\
43989 & OT2bzuckerm2 & 70 & 10.5 & 3.0 \\
&& 160 & 49.4 & 11.4  \\
54341 & OT2bzuckerm2 & 100 & 297 & 3.9 \\
&& 160 & 200 & 11  \\
76543 & OT2bzuckerm2 & 100 & 302 & 6.5 \\
&& 160 & 268 & 52 \\
76582 & OT2bzuckerm2 & 100 & 605 & 7.3 \\
&& 160 & 485 & 40.0 \\
84870 & OT2bzuckerm2 & 100 & 287 & 3.8 \\
&& 160 & 246.0 & 56.0 \\
85672 & OT2bzuckerm2 & 100 & 165 & 8.1 \\
&& 160 & 137 & 72 \\
99945 & OT2bzuckerm2 & 100 & 159 & 4.0 \\
&& 160 & 232 & 66 \\
113766 & OT1jolofsso1 &70 & 322 & 9.7 \\
&& 100 & 201 & 6.0 \\
&& 160 & 79 & 8.0 \\
121191 & OT1bzuckerm1 & 70 & 246.6 & 4.7 \\
&& 160 & 37.8 & 25.2 \\
124718 & OT2bzuckerm2 & 100 & 4.2 & 5.2 \\
&& 160 & 2.4 & 17.7 \\
131488 & OT1bzuckerm1 & 100 & 336.6 & 5.8 \\
&& 160 & 191.3 & 24.0 \\
BD+20 307 & OT1jolofsso1 & 70 & 47 & 3.0 \\
& & 100 & 16 & 3.0 \\
\hline
\multicolumn{5}{c}{Herschel Non-Detections}\\
\hline
70298 & OT2bzuckerm2 & 100 & $<$ 10.2& \\
& & 160 & $<$ 23.1& \\
72660 & OT2bzuckerm2 & 100 & $<$ 13.2& \\
& & 160 & $<$ 38.7& \\
132950 & OT2bzuckerm2 & 100 & $<$ 11.1& \\
& & 160 & $<$ 39.6& \\
%\hline
%\multicolumn{5}{c}{Contaminated Fields and Stars with Large Offsets}\\
%\hline
%8558 & OT2bzuckerm2 & 70 & 3.8 & 0.5 \\
%13183 & OT2bzuckerm2 & 70 & 2.4 & 1.6 \\
%60234 & OT2bzuckerm2 & & & \\
%80425 & OT2bzuckerm2 & 100 & 327.5 & 6.7 \\
%191692 & OT2bzuckerm2 & 100 & 16.3 & 6.3 \\
%203562 & OT2bzuckerm2 & & &
\enddata
\end{deluxetable}

\clearpage
%TABLE 3 - Dust parameters (include updated SED values)
\begin{landscape}
\begin{deluxetable}{ccccccccccccc}
\tablewidth{0pt}
\tablecaption{Dust Parameters \label{dustpars}}
\tabletypesize{\tiny}
\tablehead{
\colhead{HD}&
\colhead{T$_{d1}$}&
\colhead{T$_{d2}$}&
\colhead{R$_{BB1}$}&
\colhead{R$_{BB2}$}&
\colhead{$\tau$1$^{a}$}&
\colhead{$\tau$2$^{a}$}&
\colhead{$\tau1$$_{max}$}&
\colhead{$f1$$^{b}$}&
\colhead{M$_{d, min}$}&
\colhead{a$_{blow}$}\\
\colhead{}&
\colhead{(K)}&
\colhead{(K)}&
\colhead{(AU)}&
\colhead{(AU)}&
\colhead{}&
\colhead{}&
\colhead{}&
\colhead{}&
\colhead{(M$_{\Earth}$)}&
\colhead{($\mu$m)}}
\startdata
15407&334 $\pm$ 39&1022 $\pm$ 79&1.41 $\pm$ 0.2&0.1 $\pm$ 0.02&2.10E-03 $\pm$ 1.57E-02&9.71E-03 $\pm$ 1.57E-02&3.3E-07&6360&6.56E-07&2.5\\
23514&168 $\pm$ 7&1082 $\pm$ 22&3.53 $\pm$ 0.2&0.1 $\pm$ 0.004&1.03E-03 $\pm$ 1.50E-03&1.87E-02 $\pm$ 1.50E-03&3.54E-06&290&2.02E-06&0.98\\
35650&45 $\pm$ 3&536 $\pm$ 19&13.4 $\pm$ 1.8&0.1 $\pm$ 0.007&1.48E-04 $\pm$ 8.54E-05&2.50E-03 $\pm$ 8.54E-05&6.71E-04&0.22&4.16E-06&0.11\\
43989&112 $\pm$ 13&&7.24 $\pm$ 2.3&&7.43E-05 $\pm$ 1.19E-05&&8.54E-05&0.87&6.12E-07&1.0\\
54341&60 $\pm$ 1&168 $\pm$ 25&106.7 $\pm$ 2.9&13.6$\pm$ 4.1&2.61E-04 $\pm$ 8.88E-06&1.95E-05 $\pm$ 8.88E-06&1.64E-02&0.02&4.66E-04&8.7\\
76543&49 $\pm$ 7&105 $\pm$ 10&119.2 $\pm$ 20.2&25.9 $\pm$ 0.3&4.81E-05 $\pm$ 3.98E-06&3.64E-05 $\pm$ 3.98E-06&8.65E-04&0.06&1.07E-04&6.1\\
76582&51 $\pm$ 3&132 $\pm$ 33&89.1 $\pm$ 10.7&13.29 $\pm$ 5.5&1.71E-04 $\pm$ 1.19E-05&3.20E-05 $\pm$ 1.19E-05&7.92E-04&0.26&2.13E-04&4.5\\
84870&54 $\pm$ 2&&75.1 $\pm$ 5.4&&4.71E-04 $\pm$ 3.39E-05&&1.78E-03&0.26&4.17E-04&4.2\\
85672&77 $\pm$ 6&&37.9 $\pm$ 5.6&&5.99E-04 $\pm$ 1.03E-04&&1.11E-03&0.54&1.35E-04&4.2\\
99945&38 $\pm$ 3&&166.5 $\pm$ 27.7&&5.93E-05 $\pm$ 3.52E-06&&3.14E-03&0.02&2.58E-04&4.5\\
113766&267 $\pm$ 5&1209 $\pm$ 213&2.3 $\pm$ 0.11&0.2 $\pm$ 0.07&2.24E-02 $\pm$ 2.39E-03&1.05E-02 $\pm$ 2.39E-03&4.12E-06&5440&1.88E-05&2.0\\
121191&118 $\pm$ 2&555 $\pm$ 16&18.1 $\pm$ 0.5&0.8 $\pm$ 0.06&2.59E-03 $\pm$ 1.22E-04&2.09E-03 $\pm$ 1.22E-04&5.10E-04&5.1&1.34E-04&4.99\\
124718&179 $\pm$ 49&&2.4 $\pm$ 1.8&&2.14E-04 $\pm$ 3.05E-05&&4.16E-07&514&1.98E-07&0.9\\
131488&94 $\pm$ 1&570 $\pm$ 10&41.0 $\pm$ 0.8&1.1 $\pm$ 0.05&2.73E-03 $\pm$ 1.13E-04&2.78E-03 $\pm$ 1.13E-04&2.02E-03&1.4&7.22E-04&8.4\\
BD+20307&417 $\pm$ 14&&0.7 $\pm$ 0.05&&9.78E-03 $\pm$ 4.80E-03&&2.00E-08&490000&7.74E-07&0.9\\

\enddata
\tablecomments{$^{a}$$\tau$=L$_{IR}$/L$_{bol}$; $^{b}$ the ratio of the observed fractional IR luminosity ($\tau$) and the maximum fractional IR luminosity that can be achieved through steady state collisions ($\tau$$_{max})$. If f1 is greater than 100, then the observed dust was likely produced in a transient event. All calculated values ($\tau$$_{max}$, M$_{d, min}$, and a$_{blow}$ are derived using the outermost dust belt parameters only.}
\end{deluxetable}
\end{landscape}

\clearpage

%TABLE 4 - Resolved Disk Parameters
\begin{deluxetable}{ccccccc}
\tablewidth{0pt}
\tablecaption{Disk Parameters from Herschel Imaging \label{resolvedtable}}
\tablehead{
\colhead{Star name} & 
\colhead{$\lambda$} & 
\colhead{R$_{img}$}&
\colhead{R$_{img}$}&
\colhead{Inc.} & 
\colhead{P.A.}&
\colhead{Disk+Star Flux}*\\ 
\colhead{} & 
\colhead{($\micron$)} & 
\colhead{(AU)} & 
\colhead{(\arcsec)} & 
\colhead{($^\circ$)} & 
\colhead{($^\circ$)} &
\colhead{(mJy))}
}
\startdata
HD 54341 &100& 185 $\pm$18 & 2.0 $\pm$0.2 & 29 $\pm$20 & 72 $\pm$41&321.6 \\ 
HD 76543 &100& 162 $\pm$11 & 3.3 $\pm$0.2 & 69 $\pm$9 & 86 $\pm$7&311.1\\ 
HD 76582 &100& 216 $\pm$6 & 4.4 $\pm$0.1 & 66 $\pm$3 & 103 $\pm$3&637.5 \\ 
 &160& 235 $\pm$33 & 4.8 $\pm$0.7 & 74 $\pm$15 & 103 $\pm$12&444.4 \\ 
HD 84870 &100& 252 $\pm$16 & 2.8 $\pm$0.2 & 35 $\pm$10 & 147 $\pm$18&301.8 \\ 
HD 85672 &100& 186 $\pm$28 & 2.0 $\pm$0.3 & 28 $\pm$36 &N/A &172.8 \\ 
HD 99945 &100& 198 $\pm$18 & 3.3 $\pm$0.3 & 28 $\pm$16 & 92 $\pm$35&166.8 \\ 
HD 121191 &70&195 $\pm$26$^{1}$ & 1.5 $\pm$0.2 & 40 $\pm$17 & 25 $\pm$27&295.6 \\ 
\enddata
\tablecomments{*Best-fit disk + star flux from disk modeling (see Section \ref{resolved}). These best-fit fluxes may not always agree with the fluxes from Table 2 due to the fact that the small aperture size used in deriving the Table 2 fluxes excludes a significant amount of the extended flux due to the circumstellar disk.\\$^{1}$ See Section \ref{HD121191}.}
\end{deluxetable} 

%TABLE 5 - Summary of Results
\begin{deluxetable}{c|c|c|c|c|c|c}
\tablewidth{0pt}
\tablecaption{Summary of Results \label{summary}}
\tablehead{
\colhead{Star} & 
\colhead{Resolved?} & 
\colhead{Transient?}&
\colhead{Planet}&
\colhead{Self} &
\colhead{Two-Belt?} &
\colhead{Solid-state}\\
\colhead{Name} & 
\colhead{}& 
\colhead{}&
\colhead{Stirred?}&
\colhead{Stirred?} &
\colhead{} &
\colhead{Emission?}}
\startdata
15407&&Y&&&Y&Y\\
23514&&Y?&&&Y&Y\\
35650&&&&Y&Y&\\
43989&&&&Y&&N\\
54341&Y&&Y&&Y&N\\
76543&Y&&&Y&Y&N\\
76582&Y&&&Y&Y&N\\
84870&Y&&Y&&&N\\
85672&Y&&Y&&&\\
99945&Y&&&Y&&\\
113766&&Y&&&Y&Y\\
121191&Y&&Y&&Y&Y\\
124718&&Y?&&Y&&\\
131488&&&&Y&Y&Y\\
BD+20307&&Y&&&&Y\\
\enddata
\end{deluxetable}

\begin{figure}
\begin{center}
\includegraphics[width=0.9\columnwidth]{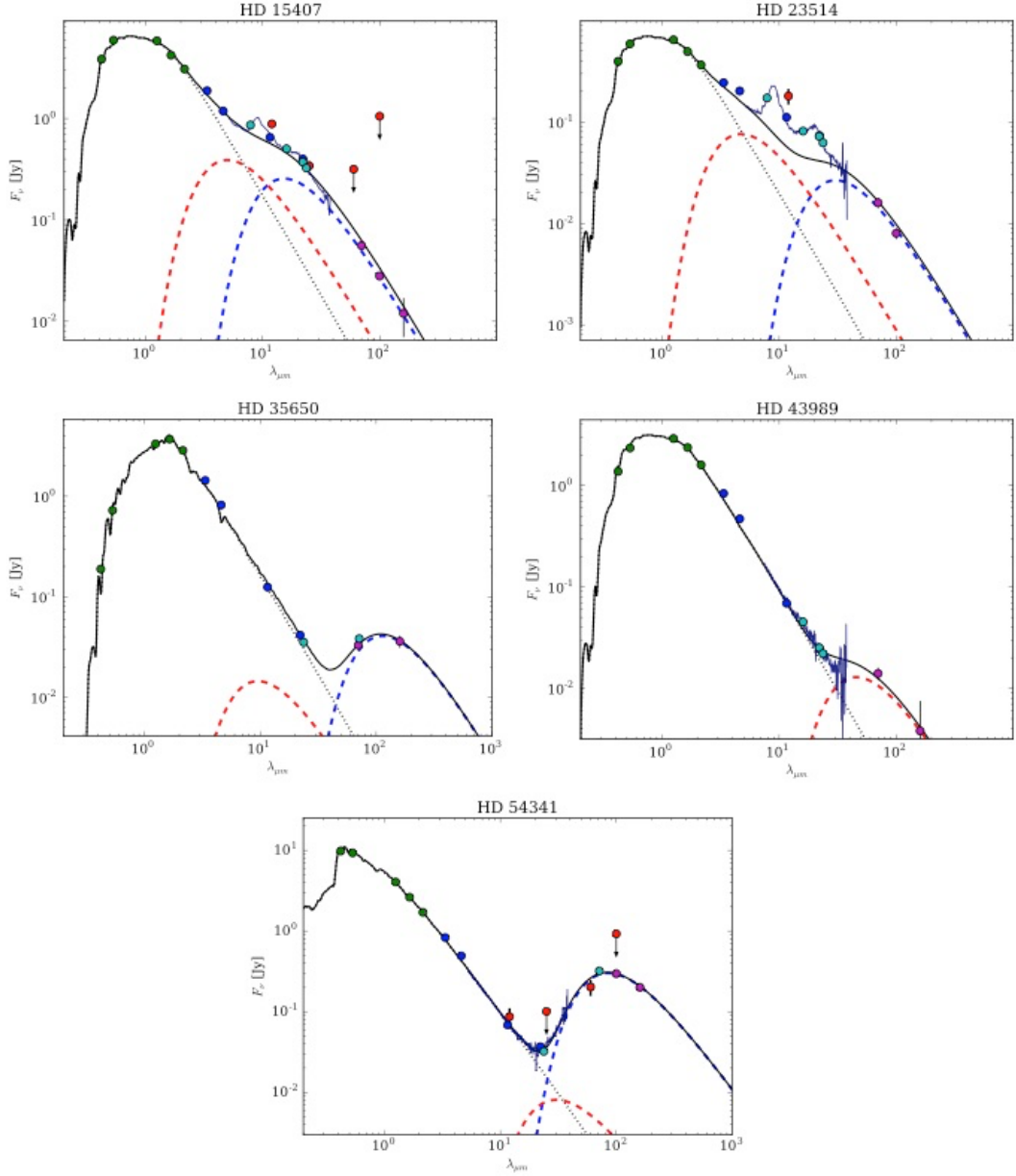}
\caption{\scriptsize Spectral energy distributions (SEDs) of OT1 and OT2 stars. Green data points are B, V, J, H, and K flux densities from the Hipparcos and 2MASS catalogs. Dark blue data points are WISE data (3.4, 4.6, 11, and 22 $\mu$m). Red data points are from IRAS. Cyan data points and blue spectra are from Spitzer. Finally, magenta data points are Herschel data points. Stellar temperature and radii were derived by fitting PHOENIX model stellar photospheres \citep{Hauschildt_1999} to photometric points. IR data is fit with one or two simple blackbodies for most targets, except in the case of HD 85672, which required a modified blackbody fit (red and blue dashed curves).}
\label{fig:SEDs1}
\end{center}
\end{figure}

\addtocounter{figure}{-1}

\begin{figure}
\begin{center}
\includegraphics[width=0.9\columnwidth]{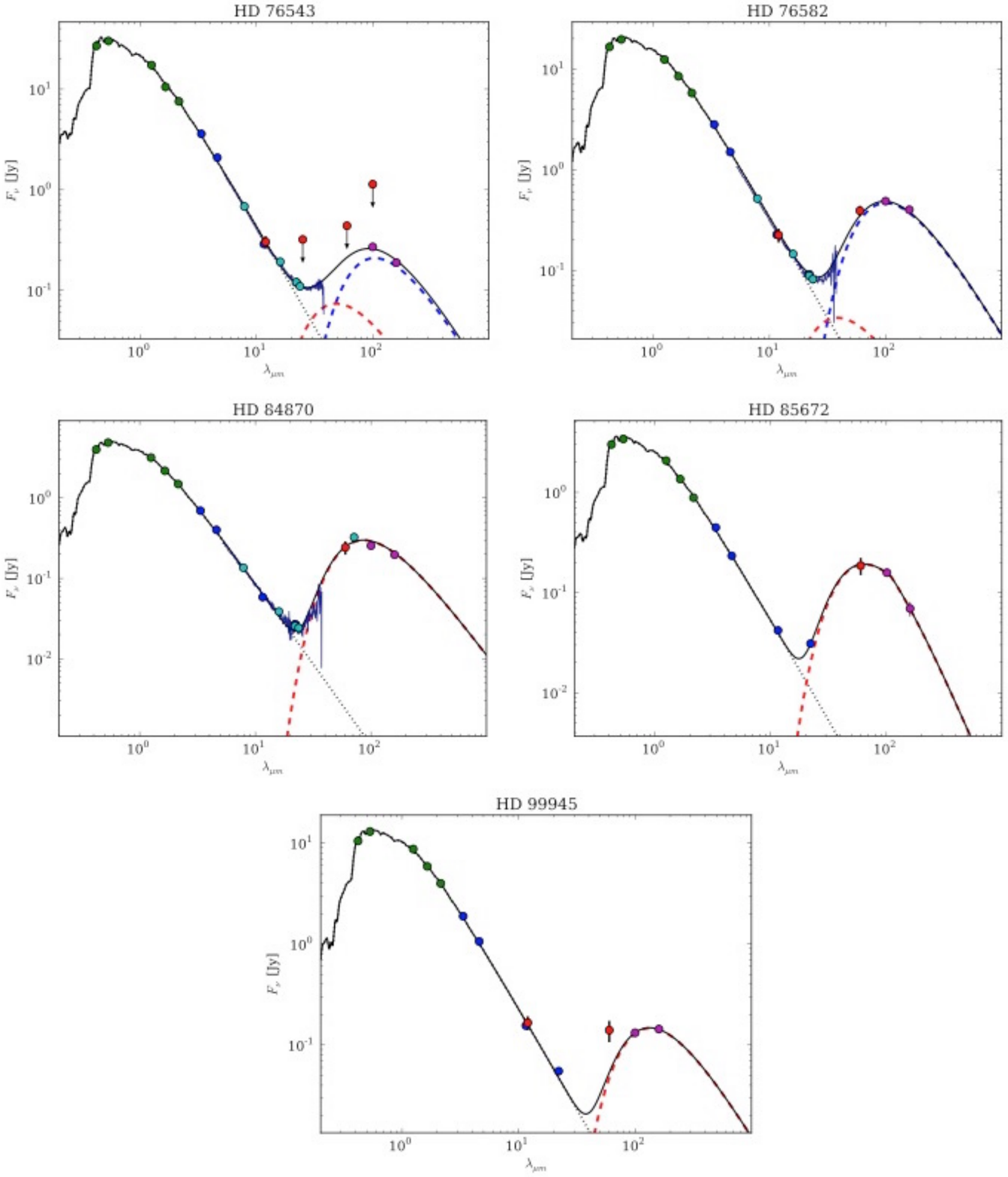}
\caption{\scriptsize Cont'd}
\label{fig:SEDs2}
\end{center}
\end{figure}

\addtocounter{figure}{-1}

\begin{figure}
\begin{center}
\includegraphics[width=0.9\columnwidth]{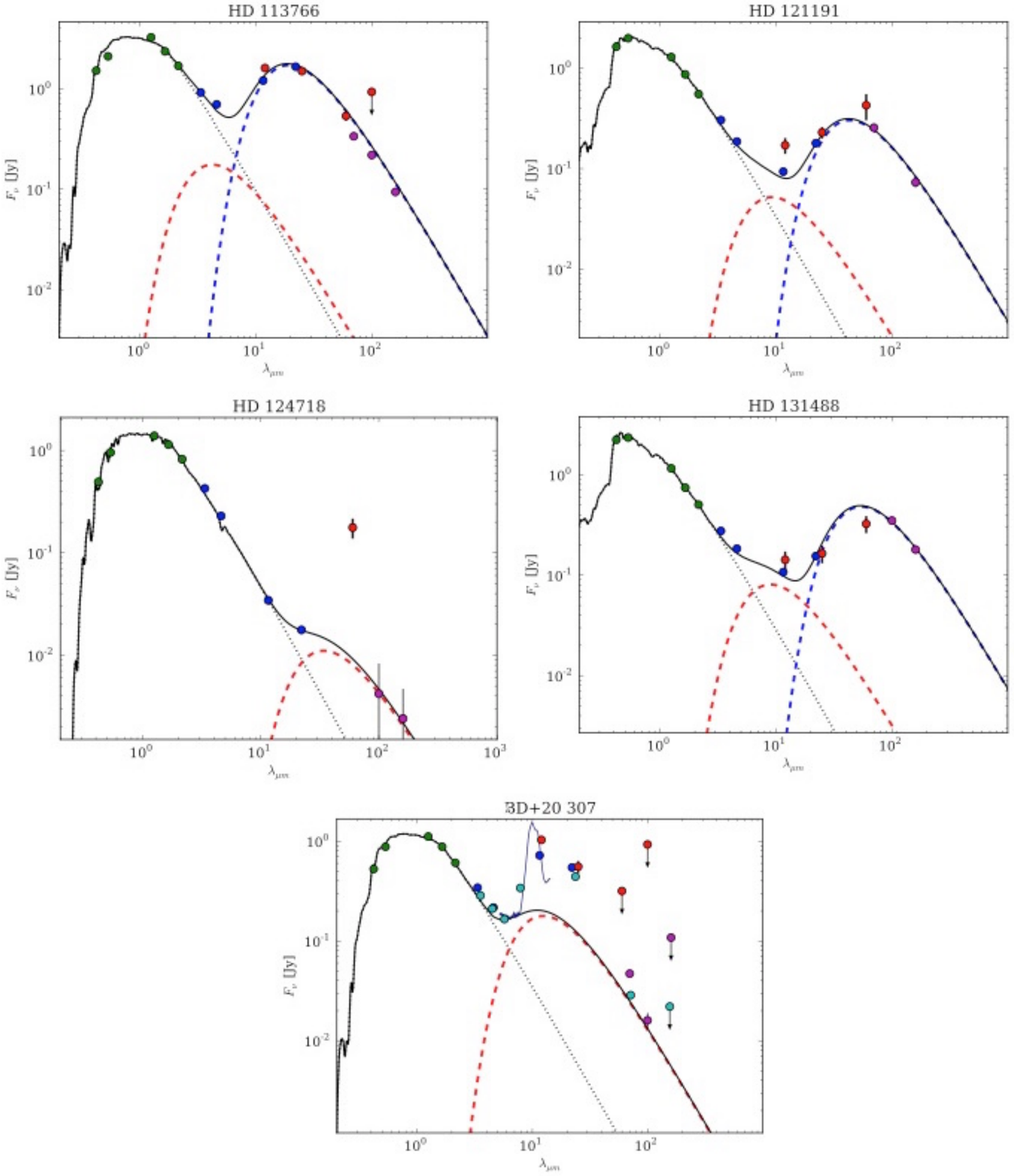}
\caption{\scriptsize Cont'd}
\label{fig:SEDs3}
\end{center}
\end{figure}

\begin{figure}[h!]
\begin{center}
\includegraphics[width=0.9\columnwidth]{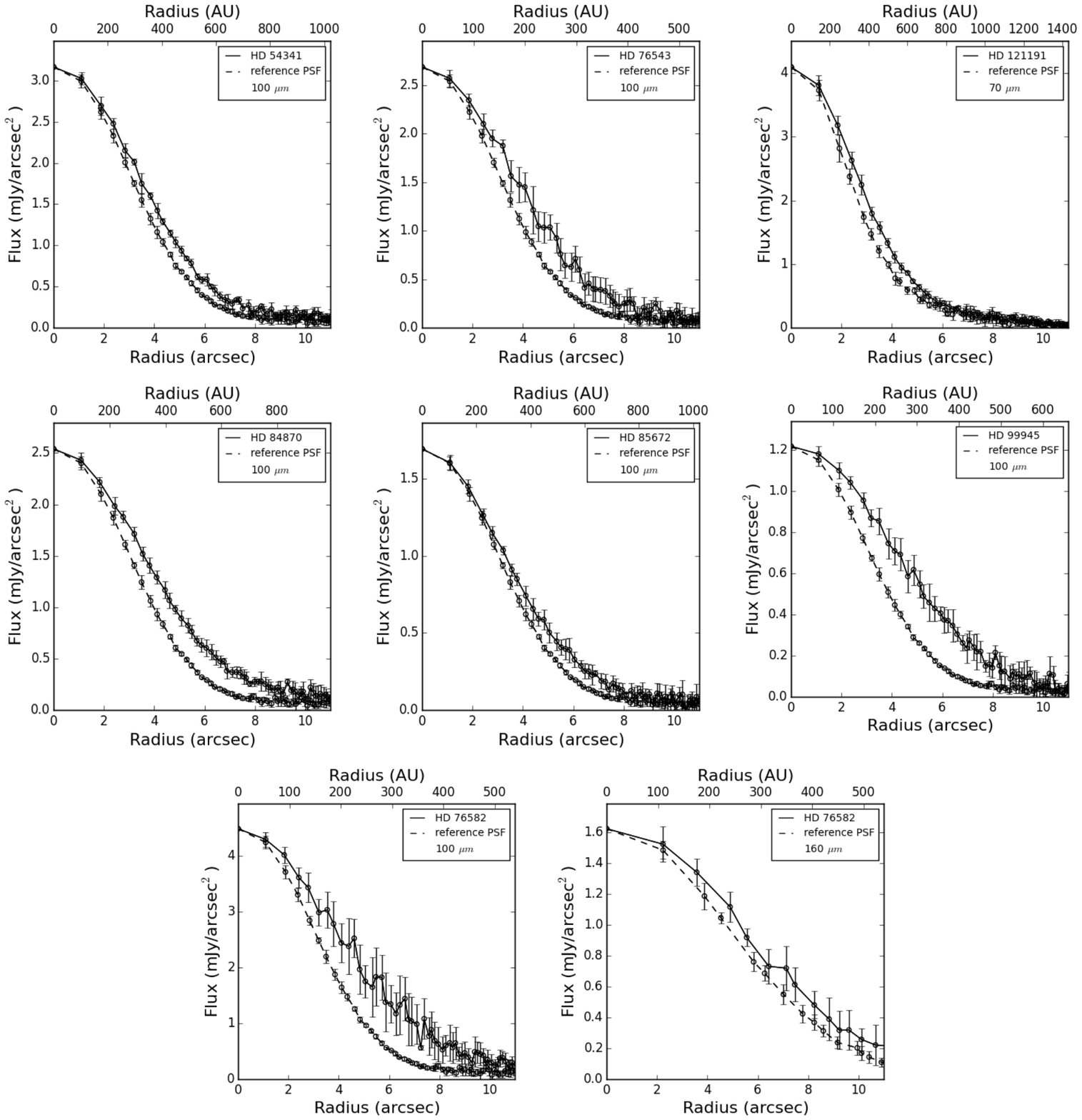}
\caption{\scriptsize Radial profiles were created to determine whether a disk was actually resolved. We compared the radial profile of the observed emission to a reference PSF (Alf Cet).}
\label{fig:profs}
\end{center}
\end{figure}

\begin{figure}[h!]
\begin{center}
\includegraphics[width=0.75\columnwidth]{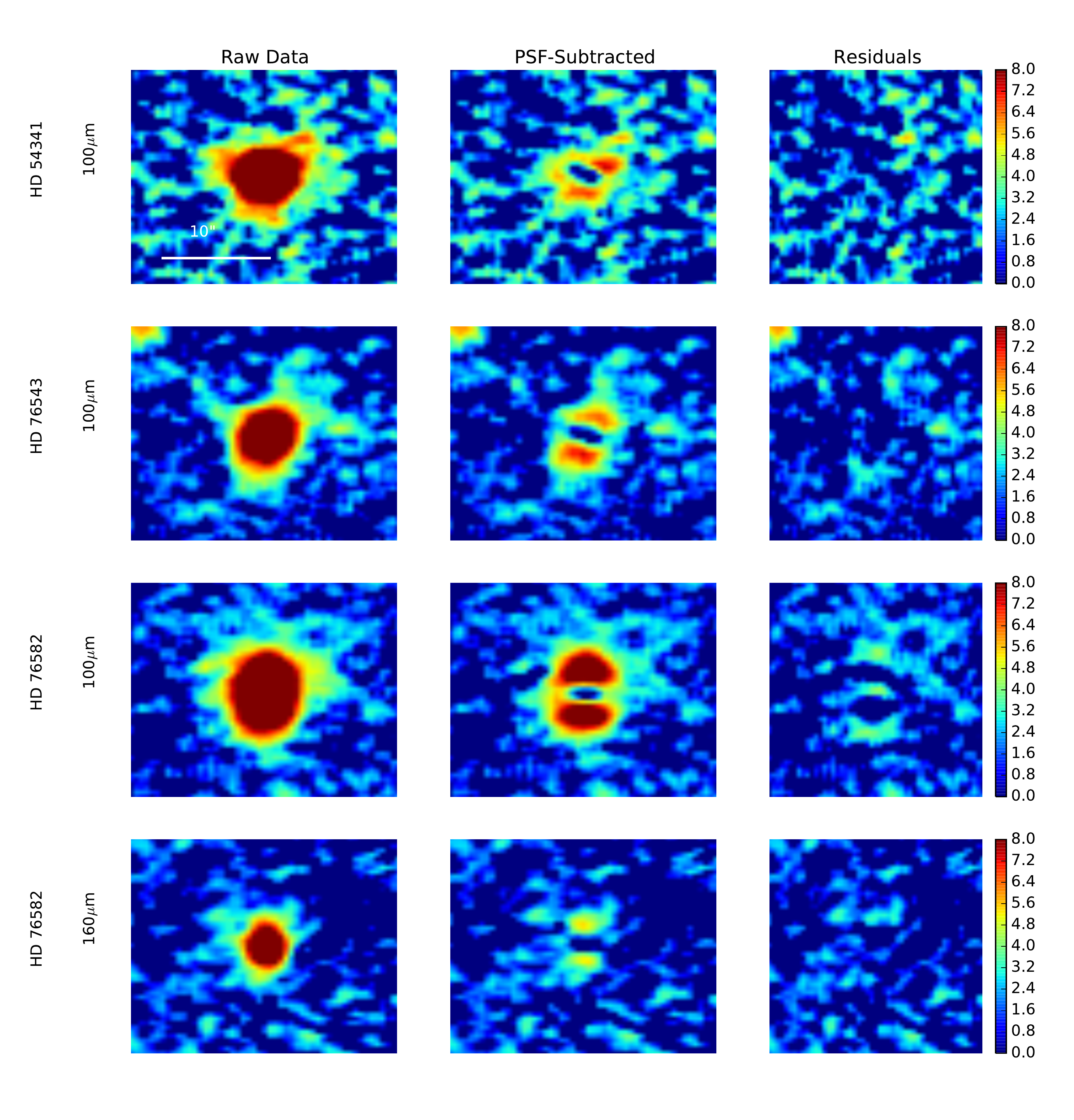}
\caption{\scriptsize Resolved disks were fit with a narrow ring of dust with four parameters: dust semi major axis, position angle, inclination, and brightness. ``PSF-subtracted" images were created by subtracting a stellar PSF which was created using a bright standard star (Alf Cet) and scaled to match the peak flux of the source. All images include the same square-root-stretch scale.}
\label{fig:mapsA}
\end{center}
\end{figure}

\addtocounter{figure}{-1}

\begin{figure}[h!]
\begin{center}
\includegraphics[width=0.75\columnwidth]{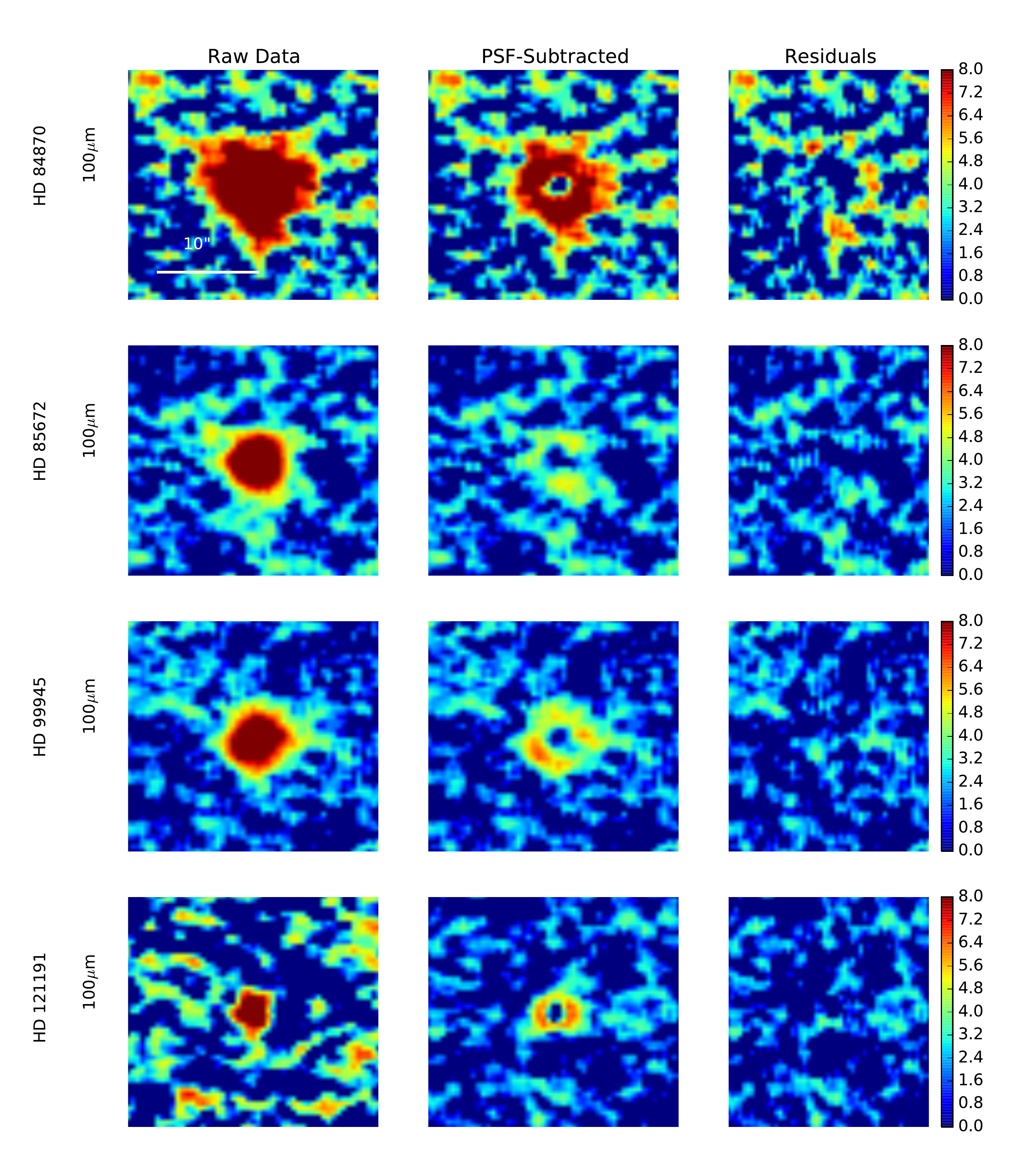}
\caption{\scriptsize Cont'd}
\label{fig:mapsB}
\end{center}
\end{figure}

\begin{figure}[h!]
\begin{center}
\includegraphics[width=0.9\columnwidth]{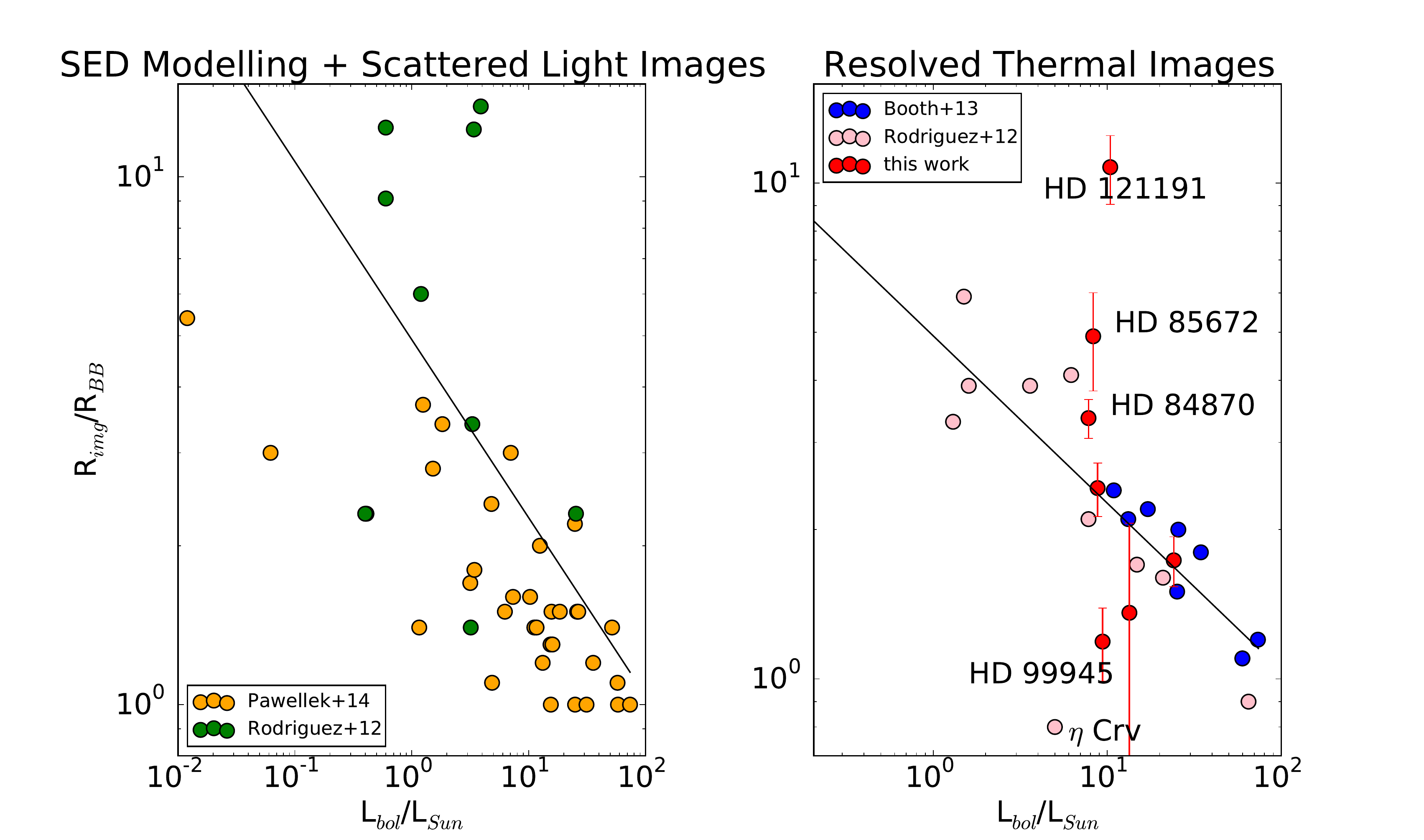}
\caption{\scriptsize Left panel: Green points represent stars compiled by \citet{Rodriguez_2012} which were observed in scattered light. Orange points represent stars in \citet{Pawellek_2014} whose ``resolved" radii are determined by emission feature - fitting in the SED. The solid black line represents a best-fit to the data from Booth et al. and Rodriguez et al. Right panel: Pink points represent stars compiled by \citet{Rodriguez_2012} which were resolved in thermal emission. Blue points represent stars from \citet{Booth_2013} which were resolved by Herschel. Red points are stars in this work that are resolved at 70 or 100 $\mu$m (see Section 6.1). The solid black line represents a best-fit to the data from Booth et al. and Rodriguez et al.. The large scatter in the red points about this best-fit line may be due to the fact that the stars in our sample with the largest discrepancies between R$_{img}$ and R$_{BB}$ are also the ones that we suspect may result from planet stirring.}
\label{fig:radii}
\end{center}
\end{figure}

\begin{figure}[h!]
\begin{center}
\includegraphics[width=0.9\columnwidth]{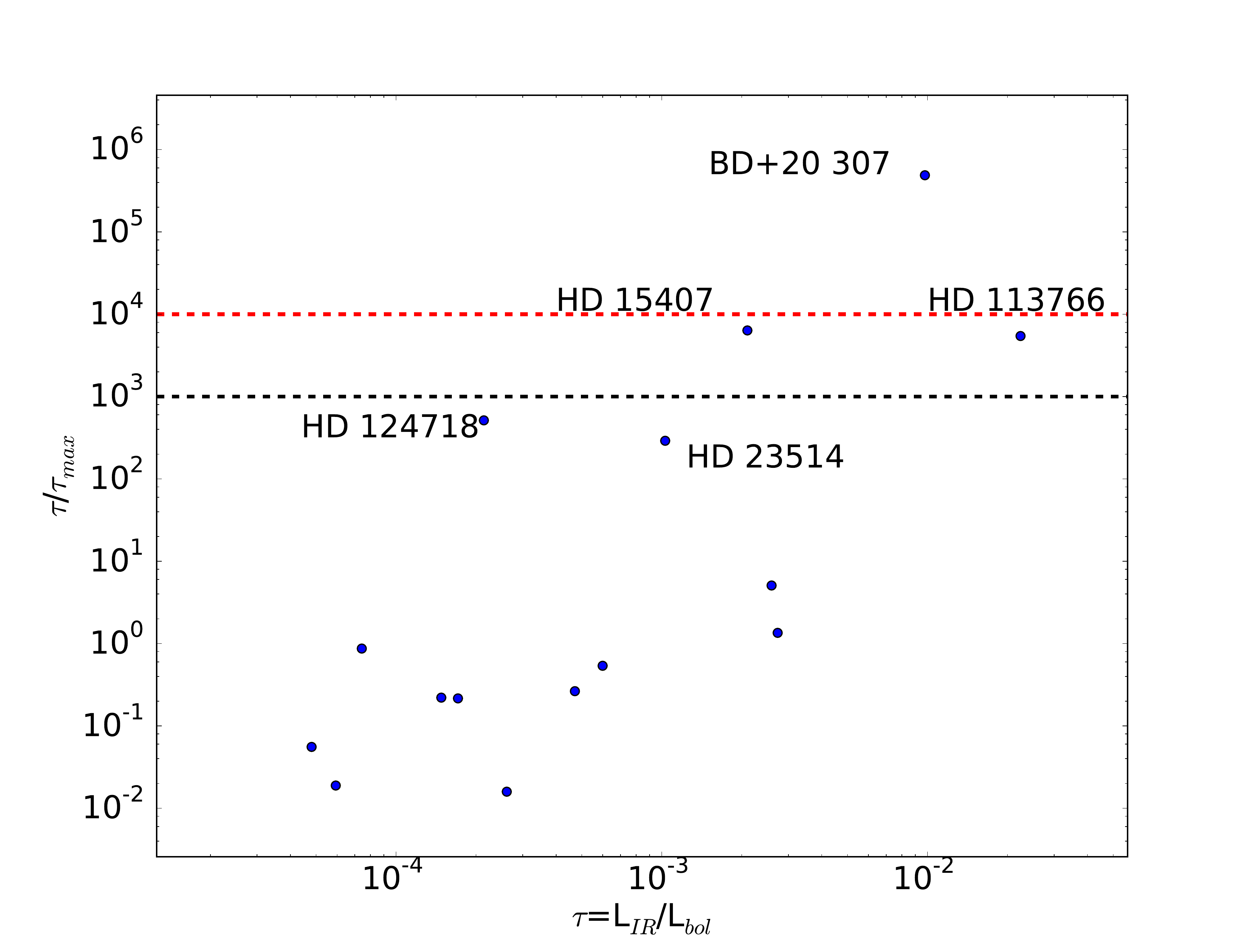}
\caption{\scriptsize Comparison of the fractional IR luminosity ($\tau$) of disks in our sample to the maximum dust luminosity achievable through steady-state collisions ($\tau$$_{max}$). If $\tau$/$\tau$$_{max}$ $>$1000 (red dashed line), the dust was likely produced in a transient event such as a catastrophic collision. Stars meeting this criterion are labeled. We also consider that disks with $\tau$/$\tau$$_{max}$ $>$100 (black dashed line) could have been created in a catastrophic collision.}
\label{fig:taumax}
\end{center}
\end{figure}

\begin{figure}[h!]
\begin{center}
\includegraphics[width=0.9\columnwidth]{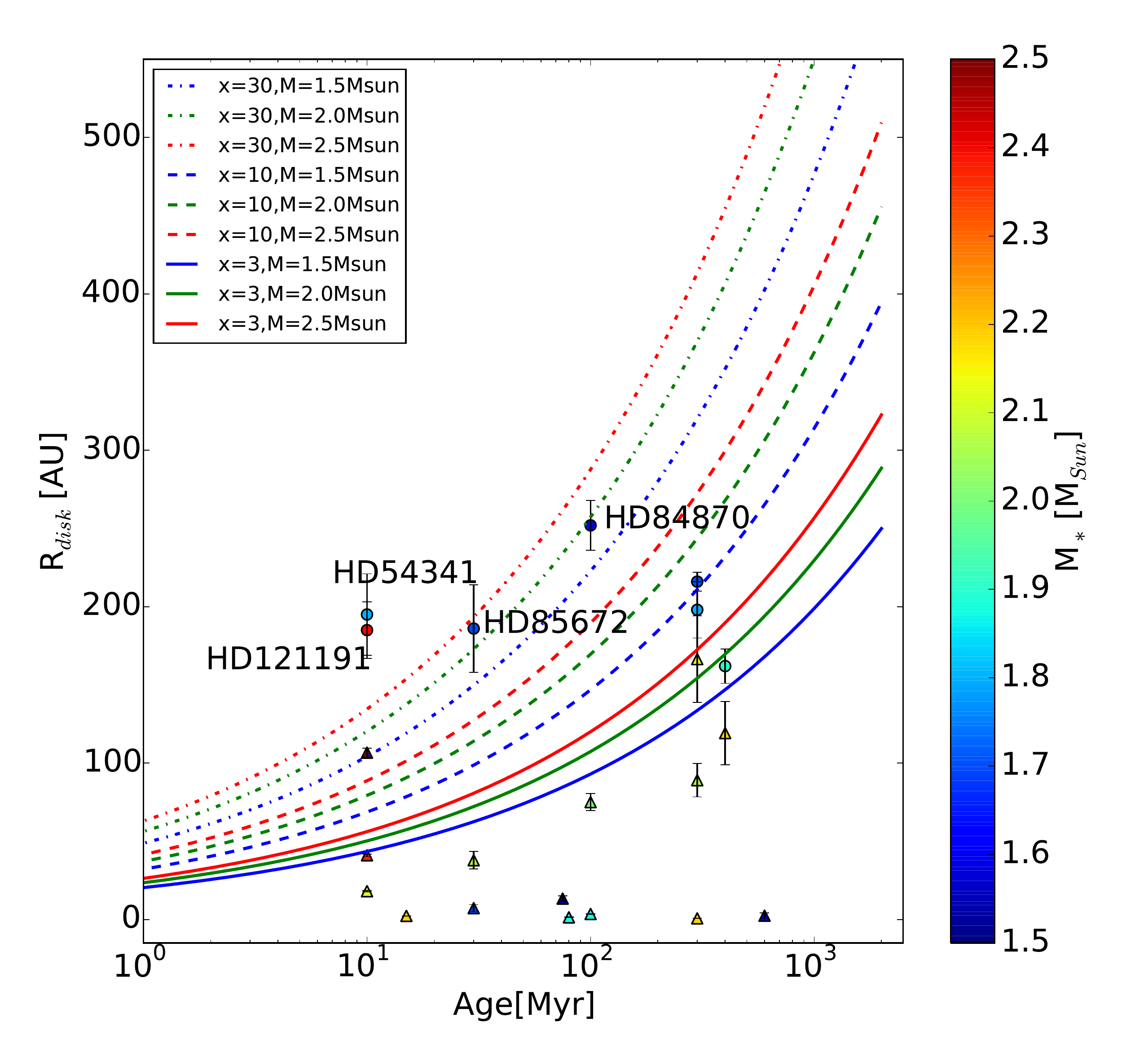}
\caption{\scriptsize Disk Radius vs. stellar age for stars in our sample. Curves represent predicted maximum radial separation of 1000km-sized bodies at a given stellar age (see Equation 4). ``x" is a scaling factor relative to the minimum mass solar nebula. Data points (circles) represent our resolved disks. Triangles represent unresolved targets for which we applied a scaling factor to the blackbody radius in order to obtain more realistic R$_{disk}$ values. See section 7.1.1 for further discussion of this figure.}
\label{fig:1000km}
\end{center}
\end{figure}

\begin{figure}[h!]
\begin{center}
\includegraphics[width=1.0\columnwidth]{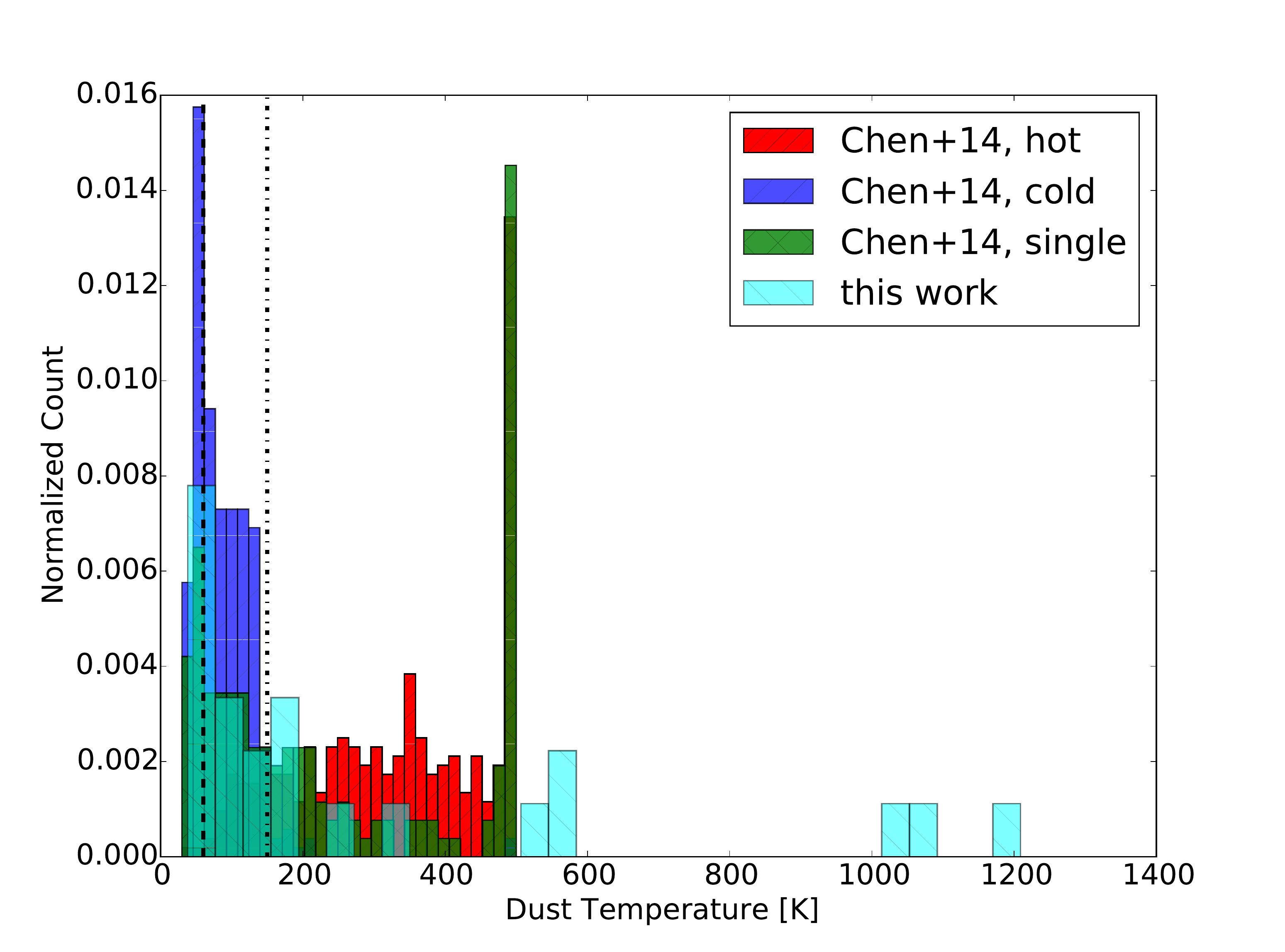}
\caption{\scriptsize Comparison of the characteristic temperatures of disks in our sample to those in the Spitzer Debris Disk Catalog \citep{Chen_2014}. We find that our disks are colder, on average, than those in the Spitzer catalog. This makes sense, since Herschel is most sensitive to colder disks. Notably, we do not find a peak in the distribution of temperatures from the disks in the Spitzer catalog that corresponds to the sublimation temperature of icy planetesimals ($\sim$150 K). We do, however, note a peak around 60 K. The build-up of disks at 500 K in the single dust belt sample is due to the fact that Chen et al. assigned a dust temperature of 500 K for anything that appeared to be 500 K or hotter.}
\label{fig:T_hist}
\end{center}
\end{figure}

\begin{figure}[h!]
\begin{center}
\includegraphics[width=0.9\columnwidth]{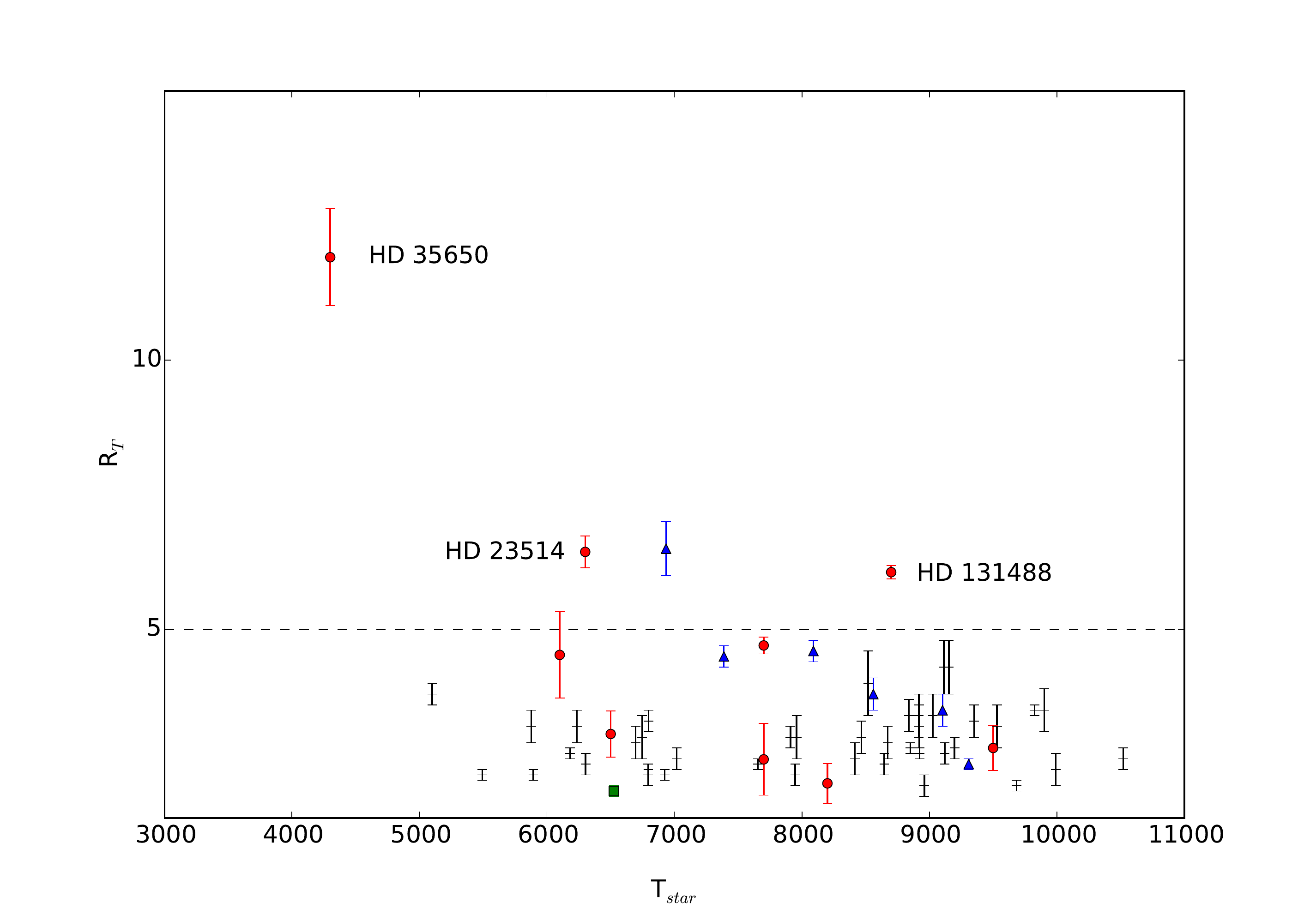}
\caption{\scriptsize The temperature ratio (R$_{T}$) between the inner and outer disks of double-belt systems can be used to distinguish spatially separated belts from single-belt systems with two grain populations \citep{Kennedy_2014}. We conservatively assume that any systems with R$_{T}$$>$5 (dashed line) are truly spatially separated systems (see Section 8.1). Colored circles represent double-belt systems from our Herschel sample (OT1 and OT2). Blue triangles represent stars from \citet{Kennedy_2014} with known (resolved) double-belt systems. Black crosses represent stars from \citet{Kennedy_2014} with double-belt fits to the dust SEDs, but for which it is not known whether there are two truly separated belts present. The green square is HD 181327, which is known to host a single belt of dust with two different temperature grain populations.}
\label{fig:doubles}
\end{center}
\end{figure}

\end{document}